\documentclass[12pt,preprint]{aastex}

\usepackage{graphicx}

\shorttitle{Understanding Compact Object Formation}
\shortauthors{B. Willems et al.}
 
\begin{document}

\title{Understanding Compact Object Formation and Natal Kicks\\ I.\
  Calculation Methods and the case of GRO\,J1655-40} 
\author{B.\ Willems$^{1}$, M.\ Henninger$^{1}$, T.\ Levin$^{1}$, N.\
  Ivanova$^{1}$, V.\ Kalogera$^{1}$, K.\ McGhee$^{2}$, F.X.\
  Timmes$^{2}$, and C.L.\ Fryer$^{2,3}$} 
\altaffiltext{1}{Northwestern University, Department of Physics and
  Astronomy, 2145 Sheridan Road, Evanston, IL 60208, USA}
\altaffiltext{2}{Theoretical Division, LANL, Los Alamos, NM 87545}
\altaffiltext{3}{Physics Dept., The University of Arizona,
  Tucson, AZ 85721}
\email{b-willems@northwestern.edu,
  m-henninger@alumni.northwestern.edu, tlevin, nata,
  vicky@northwestern.edu, kmm225@dana.ucc.nau.edu, timmes@lanl.gov,
  fryer@lanl.gov}

\begin{abstract}
In recent years proper motion measurements have been added to the set
        of observational constraints on the current properties of
        Galactic X-ray binaries. We develop an analysis that allows us
        to consider all this available information and reconstruct the
        full evolutionary history of X-ray binaries back to the time
        of core collapse and compact object formation. This analysis
        accounts for five evolutionary phases: mass transfer through
        the ongoing X-ray phase, tidal circularization before the
        onset of Roche-lobe overflow, motion through the Galactic
        potential after the formation of the compact object, binary
        orbital dynamics at the time of core collapse, and
        hydrodynamic modeling of the core collapse that connects the
        compact object to its progenitor and any nucleosynthetic
        constraints available. In this first paper, we present this
        analysis in a comprehensive manner and we apply it to the soft
        X-ray transient GRO\,J1655--40. This is the first analysis
        that incorporates all observational constraints on the current
        system properties and uses the full 3D peculiar velocity
        constraints right after core collapse instead of lower limits
        on the current space velocity given by the present-day radial
        velocity. We find that the system has remained within 200\,pc
        from the Galactic plane throughout its entire life time and
        that the mass loss and a kick possibly associated with the
        black hole formation imparted a kick velocity of $\simeq
        45$--115\,km\,s$^{-1}$ to the binary's center of mass. Right
        after black hole formation, the system consists of a $\simeq
        3.5-6.3\,M_\odot$ black hole and a $\simeq 2.3-4\,M_\odot$
        main-sequence star. At the onset of the X-ray phase the donor
        is still on the main sequence.  We find that a symmetric black
        hole formation event cannot be formally excluded, but that the
        associated system parameters are only marginally consistent
        with the currently observed binary properties. Black hole
        formation mechanisms involving an asymmetric supernova
        explosion with associated black hole kick velocities of a few
        tens of km\,s$^{-1}$, on the other hand, satisfy the
        constraints much more comfortably. We also derive an upper
        limit on the black hole kick magnitude of $\simeq 210\,{\rm
        km\,s^{-1}}$.
\end{abstract}


\section{INTRODUCTION}

The current observed sample of Galactic black-hole (BH) X-ray binaries
(XRBs) provides us with a unique opportunity for understanding the
formation of black holes in binaries. In recent years the sample has
grown significantly, reaching 18 BH XRBs identified through dynamical
mass measurements (Orosz 2004, McClintock \& Remillard 2004). For
these systems there exists a wealth of observational information about
their current physical state: BH and donor masses, orbital period,
donor's position on the H-R diagram and surface chemical composition,
transient or persistent and Roche-lobe overflow (RLO) or wind-driven
character of the mass-transfer (MT) process, and distances. Even more
recently proper motions have been measured for a handful of these
systems (Mirabel et al. 2001, 2002; Mirabel \& Rodrigues 2003a),
complementing the earlier measurements of center-of-mass radial
velocities and giving us information about the 3-dimensional kinematic
properties of these binaries. Similar information exists for some of
the known XRBs with neutron star (NS) accretors (Ankay et al. 2001;
Rib\'o et al. 2002; Mirabel \& Rodrigues 2003b; Mirabel, Rodrigues, \&
Liu 2004).

Understanding the events of compact object formation in these known
systems requires that we investigate their evolutionary history from
the present state back to the time just prior to the core collapse
event. In what follows we describe how the currently available
observational constraints for XRBs can be used to uncover this past
history and provide us with a consistent picture for how the BHs and
NSs were formed. The ultimate goal of this project is to examine the
systematics of the derived compact object progenitor masses and
requirements for supernova (SN)\footnote{Throughout the paper we use the
term SN to describe the core-collapse event leading to the formation
of both NSs and BHs. We are aware, however, that the occurrence of an
actual SN explosion during BH formation is still an unresolved
issue. The small inaccuracy in using the term SN for BHs is therefore
introduced merely for brevity. In addition, for GRO\,J1655-40 which we
consider in detail in this paper, evidence supporting a SN explosion
has been presented by Israelian et al. (1999) and Podsiadlowski et
al. (2002b).} kicks and their dependence on compact object masses
and associated mass loss at the core collapse events. Such a
comprehensive investigation will not just reveal the origin of these
Galactic XRBs, but may also offer insight to the physics of supernovae
and the possible association of natal kicks with BH formation.

Although this is not the first study that has considered the
evolutionary history of known XRBs in the Galaxy (see, e.g., de Kool,
van den Heuvel, \& Pylyser 1987; Brandt, Podsiadlowski, \& Sigurdsson
1995; Nelemans, Tauris, \& van den Heuvel 1999; Podsiadlowski,
Rappaport, \& Pfahl 2002a; Podsiadlowski, Rappaport, \& Han 2003;
Gualandris et al. 2004), it is the first one that incorporates such a
large number of observational constraints for a whole sample of
systems and considers the complete set of evolutionary processes from
the present back to the binary stage right before compact object
formation.

In this first of a series of papers, we focus on a detailed
description of our analysis methodology and the rationale for
combining the current constraints to complete a self-consistent
picture for the evolutionary history. We also focus on the
computational methods used and the uncertainties involved, and apply
our analysis to the well-studied soft X-ray transient GRO\,J1655-40
(Nova Sco 1994). In subsequent papers we will present our results for the
rest of BH XRBs and for the NS XRBs with all the above constraints
available.

The plan of the paper is as follows. In \S\,2 we briefly review the
currently available constraints on the properties of the soft X-ray
transient GRO\,J1655-40. A general outline of the analysis used to
reconstruct the system's evolutionary history is presented in \S\,3,
while the individual steps of the analysis are discussed in more
detail in \S\,4--7. In \S\,8, we combine these steps and derive
constraints on the binary's pre-SN orbital separation, on the mass of
the BH's helium star progenitor, and on the kick velocity that may
have been imparted to the BH at birth. In \S 9, hydrodynamic
core-collapse simulations are presented and the nature of the BH
progenitor is constrained even further. The final section is devoted
to a discussion of our results and some concluding remarks.

\section{OBSERVATIONAL CONSTRAINTS FOR GRO\,J1655-40}
\label{gro1655}

\clearpage
\begin{deluxetable}{lccc}
\tablecolumns{4}
\tabletypesize{\scriptsize}
\tablecaption{Properties of GRO\,J1655-40. 
\label{1655param}} 
\tablehead{ \colhead{Parameter} & 
     \colhead{Notation} & 
     \multicolumn{2}{c}{Value} 
     }
\startdata
Distance                    & $d$           & $3.2 \pm 0.2$\,kpc & (Hjellming \& Rupen 1995) \\
Galactic longitude          & $l$           & $345.0^\circ$ &  (Tingay et al. 1995) \\
Galactic latitude           & $b$           & $2.2^\circ$   &  (Tingay et al. 1995) \\
Velocity towards the Galactic Center        & $U$ & $-121 \pm 18$\,km\,s$^{-1}$ & (Mirabel et al. 2002) \\
Velocity in the direction of the Galactic Rotation    & $V$ & $-33 \pm 8$\,km\,s$^{-1}$   & (Mirabel et al. 2002) \\
Velocity towards the Northern Galactic Pole & $W$ & $3 \pm 8$\,km\,s$^{-1}$     & (Mirabel et al. 2002) \\
\cline{1-4} 
  &  & Greene et al. (2001) & Beer \& Podsiadlowski (2002) \\
\cline{1-4} 
Orbital Period              & $P_{\rm orb}$  & $2.62191 \pm 0.00020$\,days & $2.62168 \pm 0.00014$\,days\tablenotemark{a} \\
Black Hole Mass             & $M_{\rm BH}$   & $6.3 \pm 0.5\,M_\odot$  & $5.40 \pm 0.30\,M_\odot$ \\
Donor Mass                  & $M_2$          & $2.4 \pm 0.4\,M_\odot$  & $1.45 \pm 0.35\,M_\odot$ \\
Donor Luminosity            & $L_2$          & $36.5 \pm 4.5\,L_\odot$ & $21.0 \pm 6.0\,L_\odot$  \\
Donor Effective Temperature & $T_{\rm eff2}$ & $6335 \pm 350$\,K       & $6150 \pm 350$\,K         
\enddata
\tablenotetext{a}{van der Hooft et al. (1998)}
\end{deluxetable}
\clearpage 

The soft X-ray transient GRO\,J1655-40 was first detected on 27 July
1994 by the Burst and Transient Source Experiment (BATSE) on board the
Compton Gamma-Ray Observatory (Zhang et al. 1994). The optical
counterpart to the X-ray source was discovered by Bailyn et
al. (1995a). Shortly thereafter, Bailyn et al. (1995b) found periodic
optical eclipses of unequal depth which are most likely caused by
alternating eclipses of the optical component and the BH's accretion
disk. The same authors also derived the orbital period to be $2.601
\pm 0.027$\,days and used the semi-amplitude of the radial-velocity
variations to derive a minimum mass for the compact object of $3.16
\pm 0.15\,M_\odot$, strongly supporting the BH nature of the compact
star.

Since then, numerous investigations have been devoted to unraveling
the properties of the BH and its companion. Spectroscopic observations
led Orosz \& Bailyn (1997) to classify the optical component as an
F3\,IV--F6\,IV star (see also Bailyn et al. 1995b), while van der
Hooft et al. (1998) used photometric measurements to refine the
orbital period measurement to $2.62168 \pm 0.00014$\,days. The case
for the BH nature of the accreting object was strengthened
considerably by Soria et al. (1998) who derived a lower limit for the
mass of the compact star of $5.1\,M_\odot$, and by Phillips, Shahbaz
\& Podsiadlowski (1999) and Shahbaz et al. (1999) who constrained the
compact object mass to be $5.35 \pm 1.25\,M_\odot$ and $6.7 \pm
1.2\,M_\odot$, respectively. Evidence for the occurrence of a SN
explosion was presented by Israelian et al. (1999) who reported a
large overabundance of oxygen, magnesium, silicon, and sulfur in the
atmosphere of the optical component. Since the latter is not massive
enough to create these elements, Israelian et al. (1999) attributed
the overabundances to the accretion of SN material from the BH
progenitor. The distance of GRO\,J1655-40 from the Sun, finally, was
determined by Hjellming \& Rupen (1995) using a kinematic model for
the jets of the system. They derived $ d = 3.2 \pm 0.2$\,kpc, which is
in excellent agreement with previous distance estimates by McKay \&
Kesteven (1994), Tingay et al. (1995), and Bailyn et al. (1995a).

More recently Greene, Bailyn, \& Orosz (2001) and Beer \&
Podsiadlowski (2002) have undertaken detailed studies of the
ellipsoidal light variations when the system is in quiescence and
derived rather different constraints for the two component masses and
the position of the donor in the H-R diagram.

Greene et al. (2001) (hereafter GBO) used spherical NextGen atmosphere
models for a fixed effective temperature of 6336\,K (corresponding to
a spectral type of F6\,III) to model the observed B, V, I, J, K light
curves without any contribution from an accretion disk surrounding the
BH.  The model depends on two free parameters: the mass ratio $M_{\rm
BH}/M_2$, where $M_{\rm BH}$ is the mass of the BH and $M_2$ the mass
of the companion, and the orbital inclination $i$. As additional
constraints, GBO used the radial-velocity curve derived by Shahbaz et
al. (1999) and the projected rotational velocity $v_{\rm rot}\, \sin i
= 93 \pm 3 {\rm km\,s^{-1}}$ derived by Israelian et al. (1999). The
authors found a BH mass $M_{\rm BH}=6.3 \pm 0.5\,M_\odot$.

Beer \& Podsiadlowski (2002) (hereafter BP) fitted the ellipsoidal
light variations in the photometric B-, V-, R-, and I-bands using
Kurucz models for the atmosphere of the Roche-lobe filling secondary
and a BH accretion disk model with a flat temperature profile. Their
model is characterised by five free parameters: the mass ratio $M_{\rm
BH}/M_2$, the orbital inclination $i$, the polar temperature $T_{\rm
pole}$ of the donor star, the color excess $E(B-V)$, and the distance
$d$ from the Sun. They furthermore adopted the distance of $3.2 \pm
0.2$\,kpc derived by Hjellming \& Rupen (1995) as an additional ad-hoc
constraint to further tighten the limits on the derived model
parameters. In their best-fitting model, BP found $M_{\rm BH}=5.4
\pm 0.3\,M_\odot$.

At present it does not seem clear which of the two analyses (GBO or
BP) yields the most reliable measurements (see also Shahbaz 2003). In
this study we therefore consider both sets of constraints and study
them as two separate cases. An overview of all observationally
inferred properties of GRO\,J1655-40 relevant to this investigation is
presented in Table~\ref{1655param}. In particular, these properties
are the distance $d$ from the Sun, the Galactic longitude $l$ and the
Galactic latitude $b$, the center-of-mass velocity components $U$,
$V$, and $W$ with respect to a Galactic frame of reference, the
orbital period $P_{\rm orb}$, the BH mass $M_{\rm BH}$, the donor mass
$M_2$, the donor luminosity $L_2$, and the donor effective temperature
$T_{\rm eff2}$.

\section{OUTLINE OF ANALYSIS METHODOLOGY AND BASIC ASSUMPTIONS}

According to our current understanding, the formation of a BH XRB with
a low- to intermediate-mass donor star requires a primordial binary
with an extreme mass ratio and a primary (the BH progenitor) mass in
excess of $\simeq 20-25\,M_\odot$. If the period of the primordial
binary is in the range from $\simeq 1$ to $\simeq 10$\,yr, the primary
is expected to become larger than its critical Roche lobe and, due to
the extreme mass ratio, lose most of its hydrogen-rich envelope in a
dynamically unstable common-envelope phase. Provided that enough
energy is available to completely expel the envelope and avoid a
merger, the common-envelope phase results in the formation of a tight
binary consisting of a helium star (the core of the Roche-lobe filling
primary) and a relatively unevolved low- to intermediate-mass
main-sequence (MS) companion. A BH XRB is then formed when the helium star
collapses into a BH and stellar evolution and/or orbital angular
momentum losses (or exchange through tides) cause the secondary in its
turn to fill its Roche lobe and transfer mass to the BH. This
formation channel is analogous to that usually considered for the
formation of NS XRBs and has been considered previously by, e.g.,
Romani (1996), Portegies Zwart, Verbunt, \& Ergma (1997), Ergma \& van
den Heuvel (1998), Kalogera (1999), and Podsiadlowski et al. (2003).

In this paper, we restrict ourselves to the formation of BH XRBs
through the above ``standard'' evolutionary channel and do not
consider alternative formation channels involving hierarchical triple
star interactions (Eggleton \& Verbunt 1986) or tidal capture or
exchange events in dense stellar environments (Clark 1975, Hills
1976). Hence, we assume the BH's immediate progenitor to be a helium
star of mass $M_{\rm He}$ orbiting a MS
companion of mass $M_2$. In view of the strong tidal forces operating
on the binary during the common-envelope phase of the primary, we
furthermore consider the pre-SN orbit to be circular. The masses of
the BH and its companion are assumed to remain constant until the
onset of the RLO and X-ray phase.

Our goal in this analysis is to trace back the evolutionary history of
known XRBs (in this paper of the soft X-ray transient GRO\,J1655-40)
and to constrain the properties of the system's progenitor just before
and right after the SN explosion that formed the
BH\footnote{Throughout the paper we will refer to the instants just
before the SN explosion and right after the formation of the BH by the
terms pre-SN and post-SN, respectively. If the BH is formed via an
intermediary NS stage followed by fall back of some fraction
of the SN material (e.g. Brandt et al. 1995) these two times may be
slightly offset from each other. We here neglect this small offset and
assume the entire SN process to be instantaneous regardless of the
formation mechanism of the BH.}. The method adopted to derive the pre-
and post-SN constraints incorporates the following set of
calculations.

We first use a stellar evolution code and calculate a grid of
evolutionary sequences for binaries in which a BH is accreting mass
from a Roche-lobe filling companion. To consider the full range of
possibilities, we include sequences for both conservative (for
sub-Eddington rates) and fully non-conservative MT. For each sequence,
we examine whether at any point in time the calculated binary
properties, i.e., BH and donor masses, donor effective temperature and
luminosity, and orbital period, are in agreement with the
observational measurements or derivations of these quantities within
their associated uncertainties. Among these quantities, the orbital
period is measured with the highest accuracy and hence presents the
most stringent constraint to be satisfied. We consider a large number
of MT sequences and, although many of them can satisfy {\em some} of
the constraints, the majority of sequences clearly fail to
simultaneously satisfy {\em all} of the constraints at any given
time. In the case of GRO\,J1655-40 the successful sequences
furthermore have to satisfy the additional requirement that the system
exhibits transient behavior. For the time interval during which all
other constraints are satisfied, the long-term MT rate from the donor
star must therefore be lower than than the critical rate separating
transient from persistent behavior (Dubus et al. 1999; see also King,
Kolb, \& Burderi 1996; King, Kolb, \& Szuszkiewicz 1997). This last
requirement restricts the successful sequences even further. With the
remaining fully successful sequences, we derive the properties of the
binary at the onset of the RLO phase: initial BH and donor masses,
orbital period, and age of the donor star. The time at which the fully
successful sequences satisfy all observational constraints furthermore
provides an estimate for the donor's current age.

Next, we consider the kinematic evolutionary history of the XRB in the
Galactic potential. In particular, we use the current position and the
measured 3D velocity with their associated uncertainties to trace the
Galactic motion back in time. Combined with the tight constraints
on the current age of the system given by the successful MT sequences
this allows us to determine the position and velocity of the binary at
the time of BH formation (we denote these as the ``birth'' location
and velocity). By subtracting the local Galactic rotational velocity
at this position from the system's total center-of-mass velocity, we
then obtain an estimate for the {\em peculiar} velocity of the binary
right after the formation of the BH.

The ``birth'' or post-SN peculiar velocity holds information about the
mass loss and possibly the natal kick associated with the BH's
formation. In order to extract this information we must however also
constrain the orbital period and orbital eccentricity right after the
formation of the BH. We derive these constraints in the third step of
our analysis: we consider pairs of post-SN orbital periods and
eccentricities and integrate the equations governing the evolution of
the orbit under the influence of tides and general relativity
(possibly important for the most highly eccentric orbits) forward in
time. By using the age of the donor star at the onset of RLO for each
of the successful MT sequences as an estimate for the time expired
since the formation of the BH, we are able to map the post-SN orbital
parameters to those at the onset of RLO. Comparison with the binary
properties at the onset of RLO given by the sequences then allows us
to select only those pairs of post-SN orbital period and eccentricity
that can match these properties at the right time (i.e. at the right age
or evolutionary stage of the donor star).

In the fourth step of our analysis we use the derived post-SN masses,
orbital period, eccentricity, and peculiar velocity and examine the
orbital dynamics of the compact object formation allowing for a natal
kick. Based on angular momentum and energy conservation we derive
constraints on the pre-SN binary properties (BH progenitor mass and
orbital separation) and the natal kick (magnitude and direction) that
may have been imparted to the BH.

In our final step, we use the initial BH masses from the successful MT
sequences and the matching range of BH progenitor masses to study
constraints on the core collapse mechanism. By modeling the collapse
of BH progenitors and the resultant SN explosion, we can calculate the
final range of SN explosion energies and nucleosynthetic yields for
GRO\,J1655-40.  Assuming the abundance enhancements in the BH
companion are due to the SN explosion that formed the BH, we find
strong evidence for the need for asymmetries in SN explosions
associated with this BH formation event.  In turn, with current
stellar evolution and SN models, we can also further constrain the BH
progenitor mass, and ultimately the pre-SN binary properties and the
SN kick.

These five steps are described in more detail in the following
sections.

\section{MASS TRANSFER SEQUENCES}
\label{mt}

We calculate a set of evolutionary sequences for mass-transferring
binaries using an up-to-date stellar evolution code described in
detail in Podsiadlowski et al. (2002a), Ivanova et al. (2003), and
Kalogera et al. (2004). Binary orbits are assumed to be circular and
effects of stellar rotation are neglected at all times. Mass-transfer
rates are calculated self-consistently by imposing that the stellar
radius remains close to the Roche-lobe radius during the entire MT
phase. In order to determine whether or not a system is transient, we
use the criterion
\begin{eqnarray}
\dot{M} < \dot{M}_{\rm crit} & & \simeq 10^{-5}
  \left( M_{\rm BH} \over M_\odot \right)^{0.5} 
  \left( M_2 \over M_\odot\right)^{-0.2} \nonumber \\ 
 & & \times \left( P_{\rm orb} \over {\rm 1\, yr}
  \right)^{1.4} M_\odot\,\, {\rm yr^{-1}}, 
  \label{mcrit}
\end{eqnarray}
where $\dot{M}$ denotes the MT rate, and $\dot{M}_{\rm crit}$ the
critical rate separating transient from persistent behavior (Dubus et
al. 1999, Kalogera et al. 2004).  Mass accretion onto the BH is
furthermore limited to the Eddington rate at all times and any excess
mass leaving the system is assumed to carry away the specific orbital
angular momentum of the accretor. Mass-loss via stellar winds is taken
into account using the rates given in Hurley, Pols, \& Tout
(2000). For the mixing-length and convective-overshooting parameters,
finally, we adopt values of 0.2 and 0.25 pressure scale heights,
respectively.

We furthermore note that all MT sequences considered have initial
donor masses greater than $1.25\,M_\odot$ so that no magnetic braking
takes place in the donor star and the binary. For those sequences
where MT causes the donor mass to decrease below this value by just
one or two tenths of a solar mass (relevant for BP's current donor
mass estimates), magnetic braking is still not included, given the
relatively wide current orbit of the system.

Our goal with these calculations of MT sequences is to map out the
3-dimensional parameter space of BH and donor masses, and binary
orbital period {\em at the onset of RLO}, for which {\em all six}
observational constraints ({\em current} masses, orbital period, donor
luminosity and effective temperature, and transient behavior) are
satisfied {\em simultaneously}. Since each MT calculation takes a
significant amount of computational time (from a few hours to a few
days on one processor) and often requires user intervention, it is
currently not feasible to develop an automated and comprehensive
scheme for the systematic and exhaustive search of the entire
available 3-dimensional space. Instead we therefore use a
trial-and-error scheme guided strongly by the systematic behavior of
the MT sequences in {\em both} the H-R diagram and the mass-period
plane. We first examine the H-R tracks and adjust the RLO parameters
so that tracks cross the H-R error box. Next, we identify the time
intervals for which the H-R constraints are satisfied and examine
whether these parts of the tracks also cross the BH and donor mass
errors bars {\em and} the orbital period measurement, the latter being
the most stringent of all constraints. Last we examine whether the
sequence exhibits transient behavior for a significant fraction of the
time interval over which the other constraints are satisfied.

In this way we are able to outline the extent of the parameter space
that leads to ``successful'' MT sequences. It turns out that this
extent is small enough so that the variations among the successful
sequences do not significantly affect our final conclusions for
constraints on the BH XRB progenitors (see section~\ref{prog}), as
long as we stick to a given set of observational constraints (GBO or
BP). We therefore conclude that it is not necessary to strictly find
{\em all possible} successful sequences, but instead to map the
outline of their initial parameter space within about
$0.5$\,M$_{\odot}$ in the masses and about $0.1$\,d in the orbital
period.

In order to capture the uncertainties related to the accretion of
matter by the BH, we furthermore consider both conservative and
non-conservative MT sequences for each of the two sets of
observational constraints available for GRO\,J1655-40. We note that in
the conservative case, mass accretion onto the BH is still limited to
the Eddington rate and any mass transferred in excess of this rate is
assumed to be lost from the system carrying the specific orbital
angular momentum of the accretor.

\clearpage
\begin{figure*}
\resizebox{\hsize}{!}{\includegraphics{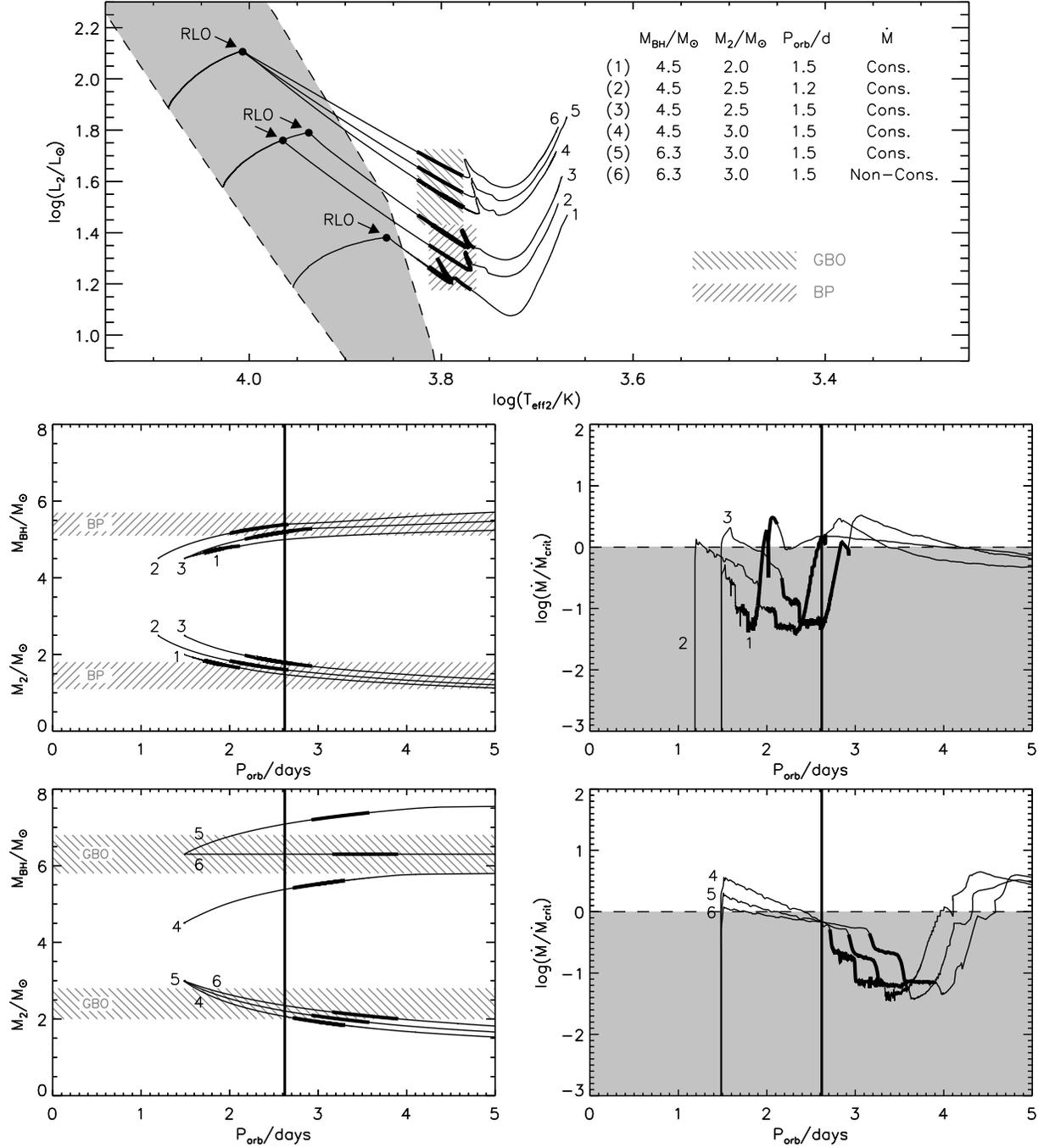}}
\caption{\scriptsize Systematic behavior of some selected MT sequences
  with different initial component masses and orbital periods. The top
  panel shows the variations of the donor star's luminosity and
  effective temperature on the H-R diagram. The MS for single stars is
  indicated by the grey-shaded region and the onset of MT by filled
  circles. The observational constraints derived by GBO and BP are
  indicated by the grey hatched regions. On the left-hand-side panels,
  the variations of the BH and donor masses are displayed as functions
  of the orbital period. The grey hatched regions indicate the
  constraints on the BH and donor mass derived by GBO and BP. On the
  right-hand-side panels the variations of the MT rate are
  displayed in units of the critical rate for transient behavior also
  as functions of the orbital period. Here, the grey-shaded region
  indicates MT rates for which the system is transient. The
  thick vertical line in the $(M,P_{\rm orb})$- and $(\log
  \dot{M}/\dot{M}_{\rm crit},P_{\rm orb})$-plots indicates the
  observed orbital period of 2.62\,days. On all five panels, the thick
  part of the evolutionary tracks indicates the part of the sequence
  where the donor star satisfies the GBO or BP constraints on the H-R
  diagram.} 
\label{tracks}
\end{figure*}
\clearpage

Before describing the sequences that are able to satisfy the
observational constraints for GRO\,J1655-40, we briefly discuss the
systematic behavior of the sequences and their dependency on the
initial component masses and orbital period. The systematic behavior
is illustrated in Fig.\,\ref{tracks} for some selected evolutionary
tracks which were chosen mainly to provide a clear and instructive
picture. The displayed sequences therefore do {\em not} represent our
best possible matches to the observed properties of GRO\,J1655-40. For
the latter, we refer to Tables~\ref{GBOseq} and~\ref{BPseq} where we
list the main properties of some selected MT sequences and where the
successful sequences are indicated in boldface. Since we did not find
any successful sequences with post-MS donor stars, we furthermore
restricted Fig.~\ref{tracks} to systems with donors that are still on
the MS.  The initial component masses and orbital period of the
sequences at the onset of MT are indicated in the top panel of the
figure. Sequences 1--5 illustrate the behavior for conservative MT,
while sequence 6 illustrates the behavior for non-conservative MT. For
convenience, the parts of the sequences where the donor star satisfies
the GBO or BP constraints in the H-R diagram are emphasized by thick
solid lines.

\clearpage
\begin{deluxetable}{lcccccccccccccccc}
\tablecolumns{17}
\tabletypesize{\scriptsize}
\rotate
\tablewidth{585pt}
\tablecaption{Selected properties of MT sequences calculated to
  satisfy the observational constraints for GRO\,J1655-40 derived by
  GBO. The current parameters correspond to the point where the
  binary's orbital period is equal to the observed orbital period of
  2.62\,days. Successful MT sequences able to satisfy the GBO
  constraints as well as the transient behavior are indicated in
  boldface.   
\label{GBOseq}}
\tablehead{ 
   \colhead{} & 
   \multicolumn{5}{c}{Parameters at onset of RLO} &
   \colhead{} &
   \multicolumn{6}{c}{Current parameters}  &
   \colhead{} &
   \multicolumn{3}{c}{SN parameters}  \\
   \cline{2-6} \cline{8-13}  \cline{15-17} \\
   \colhead{Sequence} & 
   \colhead{$M_{\rm BH}$\tablenotemark{a}} & 
   \colhead{$M_2$\tablenotemark{b}} & 
   \colhead{$P_{\rm orb}$\tablenotemark{c}} &
   \colhead{$X_2$\tablenotemark{d}} &
   \colhead{$\tau_2$\tablenotemark{e}} &
   \colhead{} &
   \colhead{$M_{\rm BH}$} & 
   \colhead{$M_2$} & 
   \colhead{$L_2$\tablenotemark{f}} &
   \colhead{$T_{\rm eff2}$\tablenotemark{g}} &
   \colhead{$X_2$} &
   \colhead{$\tau_2$} &
   \colhead{} &
   \colhead{$V_{\rm pec,postSN}$\tablenotemark{h}} &
   \colhead{$M_{\rm He}$\tablenotemark{i}} &
   \colhead{$V_{\rm k}$\tablenotemark{j}} \\
   \colhead{} & 
   \colhead{($M_\odot$)} & 
   \colhead{($M_\odot$)} & 
   \colhead{(days)} &
   \colhead{} &
   \colhead{(Myr)} &
   \colhead{} &
   \colhead{($M_\odot$)} & 
   \colhead{($M_\odot$)} & 
   \colhead{($L_\odot$)} &
   \colhead{(K)} &
   \colhead{} &
   \colhead{(Myr)} &
   \colhead{} &
   \colhead{(km\,s$^{-1}$)} &
   \colhead{($M_\odot$)} &
   \colhead{(km\,s$^{-1}$)}
   }
\startdata
\multicolumn{17}{l}{Conservative mass-transfer sequences} \\
 \\
\newcounter{GBO}
\stepcounter{GBO}GBO\arabic{GBO} & 5.5 & 2.5 & 1.5 & 0.238 & 498 & & 6.1 & 1.8 & 1.39 & 3.78 & 0.054 & 643 & & - & - & - \\
\stepcounter{GBO}GBO\arabic{GBO} & 5.5 & 2.5 & 1.7 & 0.192 & 526 & & 6.0 & 2.0 & 1.47 & 3.80 & 0.093 & 593 & & - & - & - \\
\bf \stepcounter{GBO}GBO\arabic{GBO} & \bf 5.5 & \bf 2.5 & \bf 1.8 & \bf 0.172 & \bf 537 & & \bf 6.0 & \bf 2.0 & \bf 1.51 & \bf 3.81 & \bf 0.100 & \bf 582 & & \bf 53--110 & \bf 5.5--9.5 & \bf 40--150 \\
\bf \stepcounter{GBO}GBO\arabic{GBO} & \bf 5.5 & \bf 2.5 & \bf 1.9 & \bf 0.152 & \bf 547 & & \bf 5.9 & \bf 2.1 & \bf 1.55 & \bf 3.81 & \bf 0.100 & \bf 577 & & \bf 53--110 & \bf 5.5--9.5 & \bf 40--150 \\
\bf \stepcounter{GBO}GBO\arabic{GBO} & \bf 5.5 & \bf 2.5 & \bf 2.0 & \bf 0.134 & \bf 556 & & \bf 5.9 & \bf 2.1 & \bf 1.59 & \bf 3.82 & \bf 0.097 & \bf 574 & & \bf 53--110 & \bf 5.5--9.5 & \bf 40--150 \\
\stepcounter{GBO}GBO\arabic{GBO} & 5.5 & 2.5 & 2.1 & 0.118 & 564 & & 5.8 & 2.2 & 1.62 & 3.83 & 0.088 & 575 & & - & - & - \\
%\stepcounter{GBO}GBO\arabic{GBO} & 5.5 & 3.0 & 1.0 & 0.430 & 236 & & 6.7 & 1.8 & 1.42 & 3.79 & 0.138 & 502 & & - & - & - \\
\stepcounter{GBO}GBO\arabic{GBO} & 5.5 & 3.0 & 1.2 & 0.358 & 252 & & 6.6 & 1.9 & 1.53 & 3.82 & 0.181 & 362 & & - & - & - \\
\bf \stepcounter{GBO}GBO\arabic{GBO} & \bf 5.5 & \bf 3.0 & \bf 1.3 & \bf 0.314 & \bf 303 & & \bf 6.5 & \bf 2.0 & \bf 1.59 & \bf 3.83 & \bf 0.179 & \bf 392 & & \bf 51--108 & \bf 5.5--9.5 & \bf 40--160 \\
\stepcounter{GBO}GBO\arabic{GBO} & 5.5 & 3.0 & 1.4 & 0.284 & 318 & & 6.4 & 2.1 & 1.64 & 3.84 & 0.174 & 385 & & - & - & - \\
\stepcounter{GBO}GBO\arabic{GBO} & 5.5 & 3.0 & 1.5 & 0.257 & 331 & & 6.3 & 2.2 & 1.69 & 3.85 & 0.169 & 380 & & - & - & - \\
\stepcounter{GBO}GBO\arabic{GBO} & 5.5 & 3.0 & 1.6 & 0.246 & 301 & & 6.3 & 2.2 & 1.73 & 3.85 & 0.165 & 333 & & - & - & - \\
\stepcounter{GBO}GBO\arabic{GBO} & 5.5 & 3.0 & 1.7 & 0.223 & 310 & & 6.2 & 2.3 & 1.78 & 3.86 & 0.157 & 333 & & - & - & - \\
\stepcounter{GBO}GBO\arabic{GBO} & 5.5 & 3.0 & 1.8 & 0.203 & 317 & & 6.1 & 2.4 & 1.83 & 3.87 & 0.147 & 334 & & - & - & - \\
\stepcounter{GBO}GBO\arabic{GBO} & 5.5 & 3.0 & 1.9 & 0.185 & 323 & & 5.9 & 2.5 & 1.88 & 3.88 & 0.135 & 336 & & - & - & - \\
\stepcounter{GBO}GBO\arabic{GBO} & 5.5 & 3.0 & 2.0 & 0.166 & 329 & & 5.8 & 2.5 & 1.92 & 3.89 & 0.124 & 339 & & - & - & - \\
\stepcounter{GBO}GBO\arabic{GBO} & 5.5 & 3.5 & 0.8 & 0.568 &  90 & & 7.2 & 1.8 & 1.49 & 3.81 & 0.182 & 396 & & - & - & - \\
\stepcounter{GBO}GBO\arabic{GBO} & 5.5 & 3.5 & 1.0 & 0.456 & 148 & & 7.0 & 2.0 & 1.64 & 3.84 & 0.206 & 302 & & - & - & - \\
\stepcounter{GBO}GBO\arabic{GBO} & 5.5 & 3.5 & 1.1 & 0.422 & 146 & & 6.9 & 2.1 & 1.71 & 3.85 & 0.209 & 246 & & - & - & - \\
\stepcounter{GBO}GBO\arabic{GBO} & 5.5 & 3.5 & 1.2 & 0.385 & 160 & & 6.8 & 2.2 & 1.77 & 3.86 & 0.207 & 234 & & - & - & - \\
\stepcounter{GBO}GBO\arabic{GBO} & 5.5 & 3.5 & 1.3 & 0.362 & 170 & & 6.7 & 2.3 & 1.82 & 3.88 & 0.203 & 225 & & - & - & - \\
\stepcounter{GBO}GBO\arabic{GBO} & 5.5 & 3.5 & 1.4 & 0.320 & 180 & & 6.6 & 2.3 & 1.88 & 3.89 & 0.198 & 221 & & - & - & - \\
\stepcounter{GBO}GBO\arabic{GBO} & 5.5 & 3.5 & 1.5 & 0.292 & 188 & & 6.4 & 2.4 & 1.94 & 3.90 & 0.190 & 218 & & - & - & - \\
\stepcounter{GBO}GBO\arabic{GBO} & 5.5 & 3.5 & 1.6 & 0.268 & 195 & & 6.2 & 2.6 & 2.00 & 3.91 & 0.180 & 217 & & - & - & - \\
\stepcounter{GBO}GBO\arabic{GBO} & 6.0 & 2.5 & 1.7 & 0.193 & 525 & & 6.5 & 2.0 & 1.48 & 3.80 & 0.098 & 588 & & - & - & - \\
\bf \stepcounter{GBO}GBO\arabic{GBO} & \bf 6.0 & \bf 2.5 & \bf 1.8 & \bf 0.243 & \bf 526 & & \bf 6.5 & \bf 2.0 & \bf 1.52 & \bf 3.81 & \bf 0.103 & \bf 569 & & \bf 52--105 &\bf 6.0--10.0 & \bf 40--150 \\
\bf \refstepcounter{GBO}GBO\arabic{GBO} \label{rtw} & \bf 6.0 & \bf 2.5 & \bf 1.9 & \bf 0.153 & \bf 546 & & \bf 6.4 & \bf 2.1 & \bf 1.56 & \bf 3.82 & \bf 0.101 & \bf 576 & & \bf 52--110 & \bf 6.0--10.5 & \bf 40--150 \\
\bf \stepcounter{GBO}GBO\arabic{GBO} & \bf 6.0 & \bf 2.5 & \bf 2.0 & \bf 0.127 & \bf 555 & & \bf 6.3 & \bf 2.1 & \bf 1.60 & \bf 3.82 & \bf 0.092 & \bf 574 & & \bf 53--110 & \bf 6.0--10.5 & \bf 40--150 \\
\stepcounter{GBO}GBO\arabic{GBO} & 6.0 & 2.5 & 2.1 & 0.109 & 564& & 6.3 & 2.2 & 1.63 & 3.83 & 0.085 & 576 & & - & - & - \\
\stepcounter{GBO}GBO\arabic{GBO} & 6.0 & 2.5 & 2.2 & 0.103 & 571 & & 6.2 & 2.3 & 1.67 & 3.84 & 0.076 & 580 & & - & - & - \\
\stepcounter{GBO}GBO\arabic{GBO} & 6.0 & 3.0 & 1.2 & 0.350 & 284 & & 7.0 & 2.0 & 1.56 & 3.82 & 0.175 & 407 & & - & - & - \\
\stepcounter{GBO}GBO\arabic{GBO} & 6.0 & 3.0 & 1.3 & 0.327 & 267 & & 7.0 & 2.0 & 1.61 & 3.83 & 0.180 & 345 & & - & - & - \\
\stepcounter{GBO}GBO\arabic{GBO} & 6.0 & 3.0 & 1.4 & 0.287 & 280 & & 6.9 & 2.1 & 1.66 & 3.84 & 0.173 & 339 & & - & - & - \\
\stepcounter{GBO}GBO\arabic{GBO} & 6.0 & 3.0 & 1.5 & 0.258 & 330 & & 6.8 & 2.2 & 1.71 & 3.85 & 0.165 & 381 & & - & - & - \\
\stepcounter{GBO}GBO\arabic{GBO} & 6.0 & 3.0 & 1.6 & 0.247 & 301 & & 6.7 & 2.3 & 1.75 & 3.86 & 0.161 & 335 & & - & - & - \\
\stepcounter{GBO}GBO\arabic{GBO} & 6.0 & 3.0 & 1.7 & 0.213 & 309 & & 6.7 & 2.3 & 1.80 & 3.87 & 0.152 & 334 & & - & - & - \\
\stepcounter{GBO}GBO\arabic{GBO} & 6.0 & 3.0 & 1.8 & 0.190 & 358 & & 6.6 & 2.4 & 1.84 & 3.88 & 0.143 & 379 & & - & - & - \\
\stepcounter{GBO}GBO\arabic{GBO} & 6.0 & 3.0 & 1.9 & 0.172 & 322 & & 6.5 & 2.5 & 1.88 & 3.88 & 0.134 & 336 & & - & - & - \\
\stepcounter{GBO}GBO\arabic{GBO} & 6.0 & 3.0 & 2.0 & 0.155 & 328 & & 6.4 & 2.6 & 1.93 & 3.89 & 0.123 & 338 & & - & - & - \\
\\
\multicolumn{17}{l}{Non-conservative mass-transfer sequences} \\
 \\
\stepcounter{GBO}GBO\arabic{GBO} & 6.3 & 2.2 & 1.4 & 0.244 & 712 & & 6.3 & 1.7 & 1.27 & 3.76 & 0.049 & 938 & & - & - & - \\
\stepcounter{GBO}GBO\arabic{GBO} & 6.3 & 2.2 & 1.5 & 0.257 & 727 & & 6.3 & 1.8 & 1.31 & 3.77 & 0.044 & 913 & & - & - & - \\
\stepcounter{GBO}GBO\arabic{GBO} & 6.3 & 2.2 & 1.6 & 0.193 & 756 & & 6.3 & 1.8 & 1.35 & 3.77 & 0.033 & 914 & & - & - & - \\
\stepcounter{GBO}GBO\arabic{GBO} & 6.3 & 2.2 & 1.7 & 0.171 & 774 & & 6.3 & 1.8 & 1.38 & 3.78 & 0.053 & 905 & & - & - & - \\
\stepcounter{GBO}GBO\arabic{GBO} & 6.3 & 2.2 & 1.8 & 0.201 & 781 & & 6.3 & 1.9 & 1.41 & 3.79 & 0.052 & 888 & & - & - & - \\
\stepcounter{GBO}GBO\arabic{GBO} & 6.3 & 2.2 & 1.9 & 0.131 & 803 & & 6.3 & 1.9 & 1.45 & 3.79 & 0.004 & 893 & & - & - & - \\
\stepcounter{GBO}GBO\arabic{GBO} & 6.3 & 2.2 & 2.0 & 0.117 & 803 & & 6.3 & 2.0 & 1.47 & 3.80 & 0.062 & 875 & & - & - & - \\
\stepcounter{GBO}GBO\arabic{GBO} & 6.3 & 2.2 & 2.1 & 0.142 & 818 & & 6.3 & 2.0 & 1.58 & 3.83 & 0.008 & 1201 & & - & - & - \\
\stepcounter{GBO}GBO\arabic{GBO} & 6.3 & 2.2 & 2.2 & 0.135 & 831 & & 6.3 & 2.0 & 1.64 & 3.84 & 0.008 & 1324 & & - & - & - \\
\stepcounter{GBO}GBO\arabic{GBO} & 6.3 & 2.6 & 1.0 & 0.455 & 319 & & 6.3 & 1.8 & 1.34 & 3.77 & 0.024 & 662 & & - & - & - \\
\stepcounter{GBO}GBO\arabic{GBO} & 6.3 & 2.6 & 1.2 & 0.336 & 386 & & 6.3 & 1.9 & 1.43 & 3.79 & 0.088 & 578 & & - & - & - \\
\stepcounter{GBO}GBO\arabic{GBO} & 6.3 & 2.6 & 1.3 & 0.302 & 409 & & 6.3 & 1.9 & 1.47 & 3.80 & 0.111 & 546 & & - & - & - \\
\bf \stepcounter{GBO}GBO\arabic{GBO} & \bf 6.3 & \bf 2.6 & \bf 1.4 & \bf 0.331 & \bf 417 & & \bf 6.3 & \bf 2.0 & \bf 1.51 & \bf 3.81 & \bf 0.121 & \bf 520 & & \bf 52--114 & \bf 6.5--10.5 & \bf 40--160 \\
\bf \stepcounter{GBO}GBO\arabic{GBO} & \bf 6.3 & \bf 2.6 & \bf 1.5 & \bf 0.641 & \bf 495 & & \bf 6.3 & \bf 2.1 & \bf 1.56 & \bf 3.82 & \bf 0.152 & \bf 583 & & \bf 53--110 & \bf 6.5--10.5 & \bf 40--150 \\
\bf \stepcounter{GBO}GBO\arabic{GBO} & \bf 6.3 & \bf 2.6 & \bf 1.6 & \bf 0.292 & \bf 446 & & \bf 6.3 & \bf 2.1 & \bf 1.59 & \bf 3.82 & \bf 0.124 & \bf 503 & & \bf 53--111 & \bf 6.5--11.0 & \bf 40--150 \\
\stepcounter{GBO}GBO\arabic{GBO} & 6.3 & 2.6 & 1.7 & 0.202 & 468 & & 6.3 & 2.2 & 1.62 & 3.83 & 0.114 & 5.13 & & - & - & - \\
\stepcounter{GBO}GBO\arabic{GBO} & 6.3 & 2.6 & 1.8 & 0.169 & 540 & & 6.3 & 2.3 & 1.68 & 3.84 & 0.095 & 5.82 & & - & - & - \\
%\stepcounter{GBO}GBO\arabic{GBO} & 6.3 & 2.6 & 2.0 & 0.224 & 486 & & 6.3 & 2.3 & 1.72 & 3.85 & 0.100 & 5.03 & & - & - & - \\
%\stepcounter{GBO}GBO\arabic{GBO} & 6.3 & 2.6 & 2.1 & 0.154 & 567 & & 6.3 & 2.4 & 1.76 & 3.86 & 0.118 & 5.85 & & - & - & - \\
\stepcounter{GBO}GBO\arabic{GBO} & 6.3 & 3.0 & 0.9 & 0.500 & 175 & & 6.3 & 1.9 & 1.51 & 3.81 & 0.166 & 4.06 & & - & - & - \\
\bf \stepcounter{GBO}GBO\arabic{GBO} & \bf 6.3 & \bf 3.0 & \bf 1.0 & \bf 0.433 & \bf 234 & & \bf 6.3 & \bf 2.0 & \bf 1.57 & \bf 3.82 & \bf 0.166 & \bf 429 & & \bf 53--109 & \bf 6.5--10.5 & \bf 30--160 \\
\stepcounter{GBO}GBO\arabic{GBO} & 6.3 & 3.0 & 1.1 & 0.406 & 231 & & 6.3 & 2.1 & 1.63 & 3.83 & 0.170 & 360 & & - & - & - \\
\stepcounter{GBO}GBO\arabic{GBO} & 6.3 & 3.0 & 1.2 & 0.350 & 284 & & 6.3 & 2.1 & 1.68 & 3.84 & 0.161 & 400 & & - & - & - \\
\stepcounter{GBO}GBO\arabic{GBO} & 6.3 & 3.0 & 1.3 & 0.603 & 301 & & 6.3 & 2.2 & 1.73 & 3.85 & 0.186 & 392 & & - & - & - \\
\stepcounter{GBO}GBO\arabic{GBO} & 6.3 & 3.0 & 1.4 & 0.306 & 279 & & 6.3 & 2.3 & 1.77 & 3.86 & 0.154 & 341 & & - & - & - \\
\stepcounter{GBO}GBO\arabic{GBO} & 6.3 & 3.0 & 1.5 & 0.271 & 291 & & 6.3 & 2.4 & 1.81 & 3.87 & 0.148 & 341 & & - & - & - \\
\stepcounter{GBO}GBO\arabic{GBO} & 6.3 & 3.3 & 0.7 & 0.637 &  52 & & 6.3 & 1.9 & 1.51 & 3.81 & 0.171 & 383 & & - & - & - \\
\stepcounter{GBO}GBO\arabic{GBO} & 6.3 & 3.3 & 0.8 & 0.569 &  96 & & 6.3 & 2.0 & 1.60 & 3.83 & 0.186 & 332 & & - & - & - \\
\stepcounter{GBO}GBO\arabic{GBO} & 6.3 & 3.3 & 0.9 & 0.501 & 152 & & 6.3 & 2.2 & 1.73 & 3.86 & 0.171 & 354 & & - & - & - \\
\stepcounter{GBO}GBO\arabic{GBO} & 6.3 & 3.3 & 1.0 & 0.459 & 154 & & 6.3 & 2.2 & 1.73 & 3.86 & 0.185 & 289 & & - & - & - \\
\stepcounter{GBO}GBO\arabic{GBO} & 6.3 & 3.3 & 1.1 & 0.416 & 173 & & 6.3 & 2.3 & 1.79 & 3.87 & 0.181 & 278 & & - & - & - \\
%\stepcounter{GBO}GBO\arabic{GBO} & 6.3 & 3.3 & 1.3 & 0.372 & 240 & & 6.3 & 2.4 & 1.87 & 3.88 & 0.203 & 316 & & - & - & - \\
\enddata
\tablenotetext{a}{BH mass}
\tablenotetext{b}{Donor mass}
\tablenotetext{c}{Orbital period}
\tablenotetext{d}{Central hydrogen fraction}
\tablenotetext{e}{Age}
\tablenotetext{f}{Donor luminosity}
\tablenotetext{g}{Donor effective temperature}
\tablenotetext{h}{Post-SN peculiar velocity}
\tablenotetext{i}{Mass of the BH's helium star progenitor}
\tablenotetext{j}{Kick velocity imparted to the BH}
\end{deluxetable}

\begin{deluxetable}{lcccccccccccccccc}
\tablecolumns{17}
\tabletypesize{\scriptsize}
\rotate
\tablewidth{585pt}
\tablecaption{Selected properties of mass-transfer sequences
  calculated to satisfy the observational constraints for
  GRO\,J1655-40 derived by BP. Symbols have the same meaning as in 
  Table~\ref{GBOseq}.
\label{BPseq}}
\tablehead{ 
   \colhead{} & 
   \multicolumn{5}{c}{Parameters at onset of RLO} &
   \colhead{} &
   \multicolumn{6}{c}{Current parameters} &
   \colhead{} &
   \multicolumn{3}{c}{SN parameters}  \\
   \cline{2-6} \cline{8-13}  \cline{15-17} \\
   \colhead{Sequence} & 
   \colhead{$M_{\rm BH}$} & 
   \colhead{$M_2$} & 
   \colhead{$P_{\rm orb}$} &
   \colhead{$X_2$} &
   \colhead{$\tau_2$} &
   \colhead{} &
   \colhead{$M_{\rm BH}$} & 
   \colhead{$M_2$} & 
   \colhead{$L_2$} &
   \colhead{$T_{\rm eff2}$} &
   \colhead{$X_2$} &
   \colhead{$\tau_2$} &
   \colhead{} & 
   \colhead{$V_{\rm pec,postSN}$} &
   \colhead{$M_{\rm He}$} &
   \colhead{$V_{\rm k}$} \\
   \colhead{} & 
   \colhead{($M_\odot$)} & 
   \colhead{($M_\odot$)} & 
   \colhead{(days)} &
   \colhead{} & 
   \colhead{(Myr)} &
   \colhead{} &
   \colhead{($M_\odot$)} & 
   \colhead{($M_\odot$)} & 
   \colhead{($L_\odot$)} &
   \colhead{(K)} &
   \colhead{} &
   \colhead{(Myr)} &
   \colhead{} & 
   \colhead{(km\,s$^{-1}$)} &
   \colhead{($M_\odot$)} &
   \colhead{(km\,s$^{-1}$)}
   }
\startdata
\multicolumn{17}{l}{Conservative mass-transfer sequences} \\
 \\
\newcounter{BP}
\stepcounter{BP}BP\arabic{BP} & 3.5 & 3.0 & 0.6 & 0.682 &  18 & & 5.3 & 1.2 & 1.04 & 3.73 & 0.090 & 1078 & & - & - & - \\
\stepcounter{BP}BP\arabic{BP} & 3.5 & 3.0 & 0.7 & 0.599 &  94 & & 5.2 & 1.3 & 1.13 & 3.74 & 0.052 & 902 & & - & - & - \\
\stepcounter{BP}BP\arabic{BP} & 3.5 & 3.0 & 0.8 & 0.541 & 147 & & 5.1 & 1.4 & 1.19 & 3.75 & 0.036 & 782 & & - & - & - \\
\stepcounter{BP}BP\arabic{BP} & 3.5 & 3.0 & 0.9 & 0.485 & 185 & & 5.0 & 1.5 & 1.26 & 3.77 & 0.003 & 700 & & - & - & - \\
\stepcounter{BP}BP\arabic{BP} & 3.5 & 3.0 & 1.0 & 0.429 & 215 & & 4.9 & 1.5 & 1.26 & 3.76 & 0.031 & 619 & & - & - & - \\
\stepcounter{BP}BP\arabic{BP} & 3.5 & 3.5 & 1.1 & 0.409 & 151 & & 4.9 & 1.8 & 1.49 & 3.81 & 0.216 & 278 & & - & - & - \\
\stepcounter{BP}BP\arabic{BP} & 3.5 & 3.0 & 1.2 & 0.348 & 257 & & 4.7 & 1.7 & 1.36 & 3.78 & 0.130 & 446 & & - & - & - \\
\stepcounter{BP}BP\arabic{BP} & 3.5 & 3.5 & 0.6 & 0.698 & 1 & & 5.6 & 1.3 & 1.15 & 3.75 & 0.056 & 885 & & - & - & - \\
\bf \stepcounter{BP}BP\arabic{BP} & \bf 3.5 & \bf 3.5 & \bf 0.7 & \bf 0.627 & \bf  51 & & \bf 5.5 & \bf 1.4 & \bf 1.28 & \bf 3.77 & \bf 0.004 & \bf 730 & & \bf 47--112 & \bf 4.0--6.0 & \bf 40--200 \\
\bf \stepcounter{BP}BP\arabic{BP} & \bf 3.5 & \bf 3.5 & \bf 0.8 & \bf 0.569 & \bf  87 & & \bf 5.4 & \bf 1.5 & \bf 1.28 & \bf 3.77 & \bf 0.059 & \bf 580 & & \bf 52--110 & \bf 3.5--6.0 & \bf 30--200 \\
\bf \stepcounter{BP}BP\arabic{BP} & \bf 3.5 & \bf 3.5 & \bf 0.9 & \bf 0.511 & \bf 114 & & \bf 5.3 & \bf 1.6 & \bf 1.35 & \bf 3.78 & \bf 0.128 & \bf 436 & & \bf 49--109 & \bf 3.5--6.0 & \bf 0--210 \\
\bf \stepcounter{BP}BP\arabic{BP} & \bf 3.5 & \bf 3.5 & \bf 1.0 & \bf 0.463 & \bf 134 & & \bf 5.1 & \bf 1.7 & \bf 1.42 & \bf 3.80 & \bf 0.184 & \bf 335 & & \bf 52--111 & \bf 3.5--6.5 & \bf 0--210 \\
\stepcounter{BP}BP\arabic{BP} & 3.5 & 3.5 & 1.1 & 0.409 & 151 & & 4.9 & 1.8 & 1.49 & 3.81 & 0.216 & 278 & & - & - & - \\
\stepcounter{BP}BP\arabic{BP} & 3.5 & 3.5 & 1.2 & 0.371 & 164 & & 4.7 & 1.9 & 1.56 & 3.82 & 0.229 & 241 & & - & - & - \\
\stepcounter{BP}BP\arabic{BP} & 3.5 & 4.0 & 0.6 & 0.699 & 1 & & 5.8 & 1.4 & 1.27 & 3.76 & 0.052 & 764 & & - & - & - \\
\stepcounter{BP}BP\arabic{BP} & 3.5 & 4.0 & 0.7 & 0.639 & 28 & & 5.8 & 1.5 & 1.30 & 3.77 & 0.049 & 580 & & - & - & - \\
\bf \stepcounter{BP}BP\arabic{BP} & \bf 3.5 & \bf 4.0 & \bf 0.8 & \bf 0.583 & \bf  55 & & \bf 5.5 & \bf 1.6 & \bf 1.41 & \bf 3.80 & \bf 0.157 & \bf 386 & & \bf 55--106 & \bf 4.0--6.0 & \bf 30--210 \\
\stepcounter{BP}BP\arabic{BP} & 3.5 & 4.0 & 0.9 & 0.515 & 74 & & 5.4 & 1.7 & 1.49 & 3.81 & 0.199 & 290 & & - & - & - \\
\stepcounter{BP}BP\arabic{BP} & 3.5 & 4.0 & 1.0 & 0.461 & 91 & & 5.2 & 1.8 & 1.58 & 3.83 & 0.223 & 234 & & - & - & - \\
\stepcounter{BP}BP\arabic{BP} & 3.5 & 4.0 & 1.1 & 0.419 & 102 & & 4.9 & 1.9 & 1.66 & 3.85 & 0.236 & 196 & & - & - & - \\
%\stepcounter{BP}BP\arabic{BP} & 3.5 & 4.0 & 1.3 & 0.345 & 120 & & 4.3 & 2.1 & 1.83 & 3.88 & 0.240 & 160 & & - & - & - \\
\stepcounter{BP}BP\arabic{BP} & 3.5 & 4.5 & 0.7 & 0.654 & 16 & & 5.9 & 1.6 & 1.38 & 3.79 & 0.105 & 461 & & - & - & - \\
\stepcounter{BP}BP\arabic{BP} & 3.5 & 4.5 & 0.8 & 0.588 & 37 & & 5.7 & 1.7 & 1.50 & 3.82 & 0.171 & 317 & & - & - & - \\
\stepcounter{BP}BP\arabic{BP} & 3.5 & 4.5 & 0.9 & 0.529 & 53 & & 5.5 & 1.8 & 1.61 & 3.84 & 0.199 & 245 & & - & - & - \\
\stepcounter{BP}BP\arabic{BP} & 3.5 & 4.5 & 1.0 & 0.476 & 65 & & 5.2 & 1.9 & 1.70 & 3.86 & 0.218 & 196 & & - & - & - \\
\stepcounter{BP}BP\arabic{BP} & 4.0 & 2.5 & 0.6 & 0.661 & 64 & & 5.4 & 1.1 & 0.97 & 3.71 & 0.121 & 1.320 & & - & - & - \\
\stepcounter{BP}BP\arabic{BP} & 4.0 & 2.5 & 0.7 & 0.575 & 186 & & 5.3 & 1.2 & 1.02 & 3.72 & 0.100 & 1.133 & & - & - & - \\
\stepcounter{BP}BP\arabic{BP} & 4.0 & 2.5 & 0.8 & 0.504 & 268 & & 5.2 & 1.3 & 1.07 & 3.73 & 0.071 & 1.004 & & - & - & - \\
\stepcounter{BP}BP\arabic{BP} & 4.0 & 2.5 & 0.9 & 0.486 & 323 & & 5.1 & 1.4 & 1.12 & 3.74 & 0.048 & 908 & & - & - & - \\
\stepcounter{BP}BP\arabic{BP} & 4.0 & 2.5 & 1.0 & 0.441 & 367 & & 5.0 & 1.4 & 1.15 & 3.74 & 0.038 & 844 & & - & - & - \\
\stepcounter{BP}BP\arabic{BP} & 4.0 & 2.5 & 1.1 & 0.358 & 411 & & 5.0 & 1.5 & 1.19 & 3.75 & 0.000 & 806 & & - & - & - \\
\stepcounter{BP}BP\arabic{BP} & 4.0 & 2.5 & 1.2 & 0.317 & 437 & & 4.9 & 1.5 & 1.24 & 3.76 & 0.046 & 768 & & - & - & - \\
\stepcounter{BP}BP\arabic{BP} & 4.0 & 2.5 & 1.3 & 0.288 & 464 & & 4.9 & 1.6 & 1.32 & 3.77 & 0.004 & 739 & & - & - & - \\
\stepcounter{BP}BP\arabic{BP} & 4.0 & 3.0 & 0.6 & 0.688 & 12 & & 5.7 & 1.3 & 1.09 & 3.73 & 0.063 & 989 & & - & - & - \\
\stepcounter{BP}BP\arabic{BP} & 4.0 & 3.0 & 0.7 & 0.604 & 90 & & 5.6 & 1.4 & 1.17 & 3.75 & 0.049 & 836 & & - & - & - \\
\bf \refstepcounter{BP}BP\arabic{BP} \label{bpeg} & \bf 4.0 & \bf 3.0 & \bf 0.8 & \bf 0.535 & \bf 144 & & \bf 5.5 & \bf 1.5 & \bf 1.27 & \bf 3.77 & \bf 0.001 & \bf 733 & & \bf 51--112 & \bf 4.0--7.0 & \bf 30--190 \\
\bf \stepcounter{BP}BP\arabic{BP} & \bf 4.0 & \bf 3.0 & \bf 0.9 & \bf 0.489 & \bf 182 & & \bf 5.5 & \bf 1.5 & \bf 1.26 & \bf 3.76 & \bf 0.016 & \bf 648 & & \bf 49--115 & \bf 4.0--7.0 & \bf 0--200 \\
\bf \stepcounter{BP}BP\arabic{BP} & \bf 4.0 & \bf 3.0 & \bf 1.0 & \bf 0.450 & \bf 244 & & \bf 5.4 & \bf 1.6 & \bf 1.31 & \bf 3.77 & \bf 0.091 & \bf 618 & & \bf 48--115 & \bf 4.0--7.0 & \bf 0--200 \\
\bf \stepcounter{BP}BP\arabic{BP} & \bf 4.0 & \bf 3.0 & \bf 1.1 & \bf 0.398 & \bf 235 & & \bf 5.3 & \bf 1.7 & \bf 1.36 & \bf 3.78 & \bf 0.117 & \bf 464 & & \bf 51--110 & \bf 4.0--7.0 & \bf 20--190 \\
\bf \stepcounter{BP}BP\arabic{BP} & \bf 4.0 & \bf 3.0 & \bf 1.2 & \bf 0.351 & \bf 255 & & \bf 5.2 & \bf 1.8 & \bf 1.42 & \bf 3.79 & \bf 0.156 & \bf 407 & & \bf 55--113 & \bf 4.0--7.5 & \bf 30--200 \\
\stepcounter{BP}BP\arabic{BP} & 4.0 & 3.0 & 1.3 & 0.309 & 270 & & 5.1 & 1.8 & 1.47 & 3.80 & 0.176 & 365 & & - & - & - \\
\stepcounter{BP}BP\arabic{BP} & 4.0 & 3.0 & 1.4 & 0.279 & 283 & & 5.0 & 1.9 & 1.53 & 3.81 & 0.185 & 343 & & - & - & - \\
\stepcounter{BP}BP\arabic{BP} & 4.0 & 3.0 & 1.5 & 0.251 & 333 & & 4.9 & 2.0 & 1.59 & 3.82 & 0.185 & 375 & & - & - & - \\
\stepcounter{BP}BP\arabic{BP} & 4.0 & 3.5 & 0.6 & 0.699 & 1 & & 6.0 & 1.4 & 1.24 & 3.75 & 0.058 & 791 & & - & - & - \\
\stepcounter{BP}BP\arabic{BP} & 4.0 & 3.5 & 0.7 & 0.626 & 47 & & 6.0 & 1.5 & 1.27 & 3.77 & 0.016 & 651 & & - & - & - \\
\stepcounter{BP}BP\arabic{BP} & 4.0 & 3.5 & 0.8 & 0.566 & 85 & & 5.9 & 1.6 & 1.34 & 3.78 & 0.111 & 480 & & - & - & - \\
\stepcounter{BP}BP\arabic{BP} & 4.0 & 3.5 & 0.9 & 0.499 & 111 & & 5.8 & 1.7 & 1.42 & 3.80 & 0.170 & 366 & & - & - & - \\
\stepcounter{BP}BP\arabic{BP} & 4.0 & 3.5 & 1.0 & 0.468 & 132 & & 5.6 & 1.8 & 1.49 & 3.81 & 0.208 & 298 & & - & - & - \\
\stepcounter{BP}BP\arabic{BP} & 4.0 & 3.5 & 1.1 & 0.401 & 149 & & 5.5 & 1.9 & 1.56 & 3.82 & 0.218 & 260 & & - & - & - \\
\stepcounter{BP}BP\arabic{BP} & 4.0 & 3.5 & 1.2 & 0.375 & 163 & & 5.3 & 2.0 & 1.63 & 3.84 & 0.223 & 238 & & - & - & - \\
\stepcounter{BP}BP\arabic{BP} & 4.0 & 3.5 & 1.3 & 0.341 & 174 & & 5.1 & 2.1 & 1.70 & 3.85 & 0.220 & 224 & & - & - & - \\
%\stepcounter{BP}BP\arabic{BP} & 4.0 & 4.0 & 0.7 & 0.654 & 26 & & 6.3 & 1.6 & 1.38 & 3.79 & 0.130 & 443 & & - & - & - \\
%\stepcounter{BP}BP\arabic{BP} & 4.0 & 4.0 & 0.9 & 0.531 & 74 & & 6.0 & 1.8 & 1.59 & 3.83 & 0.218 & 252 & & - & - & - \\
\stepcounter{BP}BP\arabic{BP} & 4.5 & 2.0 & 1.2 & 0.280 & 880 & & 5.2 & 1.3 & 1.03 & 3.72 & 0.094 & 1271 & & - & - & - \\
\stepcounter{BP}BP\arabic{BP} & 4.5 & 2.0 & 1.3 & 0.245 & 915 & & 5.1 & 1.4 & 1.05 & 3.72 & 0.097 & 1277 & & - & - & - \\
\stepcounter{BP}BP\arabic{BP} & 4.5 & 2.0 & 1.4 & 0.217 & 959 & & 5.1 & 1.4 & 1.06 & 3.72 & 0.091 & 1219 & & - & - & - \\
\stepcounter{BP}BP\arabic{BP} & 4.5 & 2.0 & 1.5 & 0.224 & 982 & & 5.0 & 1.5 & 1.08 & 3.72 & 0.088 & 1194 & & - & - & - \\
\stepcounter{BP}BP\arabic{BP} & 4.5 & 2.0 & 1.6 & 0.164 & 1016 & & 4.9 & 1.5 & 1.10 & 3.72 & 0.085 & 1188 & & - & - & - \\
\stepcounter{BP}BP\arabic{BP} & 4.5 & 2.0 & 1.7 & 0.138 & 1.032 & & 4.8 & 1.6 & 1.33 & 3.76 & 0.062 & 1193 & & - & - & - \\
\stepcounter{BP}BP\arabic{BP} & 4.5 & 2.0 & 1.8 & 0.114 & 1.054 & & 4.8 & 1.6 & 1.14 & 3.73 & 0.083 & 1752 & & - & - & - \\
\stepcounter{BP}BP\arabic{BP} & 4.5 & 2.0 & 1.9 & 0.091 & 1.075 & & 4.8 & 1.7 & 1.17 & 3.73 & 0.078 & 2376 & & - & - & - \\
\stepcounter{BP}BP\arabic{BP} & 4.5 & 2.5 & 1.1 & 0.361 & 409 & & 5.4 & 1.5 & 1.22 & 3.75 & 0.000 & 790 & & - & - & - \\
\stepcounter{BP}BP\arabic{BP} & 4.5 & 2.5 & 1.2 & 0.323 & 438 & & 5.4 & 1.6 & 1.32 & 3.77 & 0.001 & 755 & & - & - & - \\
\bf \stepcounter{BP}BP\arabic{BP} & \bf 4.5 & \bf 2.5 & \bf 1.3 & \bf 0.341 & \bf 456 & & \bf 5.3 & \bf 1.7 & \bf 1.32 & \bf 3.77 & \bf 0.003 & \bf 715 & & \bf 47--115 & \bf 4.5--7.5 & \bf 40--170 \\
\bf \refstepcounter{BP}BP\arabic{BP} \label{example1} & \bf 4.5 & \bf 2.5 & \bf 1.4 & \bf 0.315 & \bf 476 & & \bf 5.3 & \bf 1.7 & \bf 1.33 & \bf 3.77 & \bf 0.009 & \bf 689 & & \bf 52--110 & \bf 4.5--8.0 & \bf 40--170 \\
\bf \stepcounter{BP}BP\arabic{BP} & \bf 4.5 & \bf 2.5 & \bf 1.5 & \bf 0.295 & \bf 491 & & \bf 5.2 & \bf 1.8 & \bf 1.34 & \bf 3.77 & \bf 0.029 & \bf 657 & & \bf 50--114 & \bf 4.5--8.0 & \bf 40--170 \\
\stepcounter{BP}BP\arabic{BP} & 4.5 & 2.5 & 1.6 & 0.273 & 506 & & 5.1 & 1.8 & 1.39 & 3.78 & 0.053 & 623 & & - & - & - \\
\stepcounter{BP}BP\arabic{BP} & 4.5 & 3.0 & 0.8 & 0.544 & 141 & & 6.0 & 1.5 & 1.27 & 3.77 & 0.008 & 695 & & - & - & - \\
\stepcounter{BP}BP\arabic{BP} & 4.5 & 3.0 & 0.9 & 0.479 & 180 & & 5.9 & 1.6 & 1.29 & 3.77 & 0.046 & 593 & & - & - & - \\
\stepcounter{BP}BP\arabic{BP} & 4.5 & 3.0 & 1.0 & 0.436 & 211 & & 5.8 & 1.7 & 1.35 & 3.78 & 0.106 & 496 & & - & - & - \\
\stepcounter{BP}BP\arabic{BP} & 4.5 & 3.0 & 1.1 & 0.392 & 235 & & 5.7 & 1.8 & 1.41 & 3.79 & 0.145 & 429 & & - & - & - \\
\stepcounter{BP}BP\arabic{BP} & 4.5 & 3.0 & 1.2 & 0.354 & 254 & & 5.7 & 1.8 & 1.46 & 3.80 & 0.175 & 380 & & - & - & - \\
\stepcounter{BP}BP\arabic{BP} & 4.5 & 3.0 & 1.3 & 0.322 & 269 & & 5.6 & 1.9 & 1.52 & 3.81 & 0.184 & 356 & & - & - & - \\
\stepcounter{BP}BP\arabic{BP} & 4.5 & 3.0 & 1.4 & 0.292 & 282 & & 5.5 & 2.0 & 1.57 & 3.82 & 0.188 & 339 & & - & - & - \\
\stepcounter{BP}BP\arabic{BP} & 4.5 & 3.0 & 1.5 & 0.276 & 292 & & 5.4 & 2.1 & 1.63 & 3.83 & 0.182 & 331 & & - & - & - \\
\\
\multicolumn{17}{l}{Non-conservative mass-transfer sequences} \\
 \\
\stepcounter{BP}BP\arabic{BP} & 5.4 & 2.0 & 1.0 & 0.361 & 760 & & 5.4 & 1.4 & 1.03 & 3.72 & 0.099 & 1315 & & - & - & - \\
\stepcounter{BP}BP\arabic{BP} & 5.4 & 2.0 & 1.1 & 0.341 & 821 & & 5.4 & 1.4 & 1.05 & 3.72 & 0.095 & 1267 & & - & - & - \\
\stepcounter{BP}BP\arabic{BP} & 5.4 & 2.0 & 1.2 & 0.304 & 872 & & 5.4 & 1.5 & 1.07 & 3.72 & 0.092 & 1236 & & - & - & - \\
\stepcounter{BP}BP\arabic{BP} & 5.4 & 2.0 & 1.3 & 0.275 & 915 & & 5.4 & 1.5 & 1.09 & 3.72 & 0.086 & 1215 & & - & - & - \\
\stepcounter{BP}BP\arabic{BP} & 5.4 & 2.0 & 1.4 & 0.213 & 1079 & & 5.4 & 1.6 & 1.12 & 3.72 & 0.081 & 1357 & & - & - & - \\
\stepcounter{BP}BP\arabic{BP} & 5.4 & 2.0 & 1.5 & 0.208 & 1138 & & 5.4 & 1.6 & 1.17 & 3.73 & 0.062 & 1411 & & - & - & - \\
\stepcounter{BP}BP\arabic{BP} & 5.4 & 2.3 & 0.8 & 0.498 & 349 & & 5.4 & 1.4 & 1.10 & 3.73 & 0.066 & 1044 & & - & - & - \\
\stepcounter{BP}BP\arabic{BP} & 5.4 & 2.3 & 0.9 & 0.440 & 424 & & 5.4 & 1.5 & 1.13 & 3.73 & 0.054 & 978 & & - & - & - \\
\stepcounter{BP}BP\arabic{BP} & 5.4 & 2.3 & 1.0 & 0.383 & 547 & & 5.4 & 1.6 & 1.25 & 3.76 & 0.000 & 1035 & & - & - & - \\
\bf \stepcounter{BP}BP\arabic{BP} & \bf 5.4 & \bf 2.3 & \bf 1.1 & \bf 0.341 & \bf 596 & & \bf 5.4 & \bf 1.7 & \bf 1.29 & \bf 3.76 & \bf 0.000 & \bf 998 & & \bf 46--111 & \bf 5.5--9.0 & \bf 40--160  \\
\stepcounter{BP}BP\arabic{BP} & 5.4 & 2.3 & 1.2 & 0.348 & 557 & & 5.4 & 1.7 & 1.25 & 3.76 & 0.091 & 860 & & - & - & - \\
\stepcounter{BP}BP\arabic{BP} & 5.4 & 2.5 & 0.8 & 0.543 & 257 & & 5.4 & 1.5 & 1.20 & 3.75 & 0.042 & 854 & & - & - & - \\
\bf \stepcounter{BP}BP\arabic{BP} & \bf 5.4 & \bf 2.5 & \bf 0.9 & \bf 0.445 & \bf 389 & & \bf 5.4 & \bf 1.6 & \bf 1.33 & \bf 3.77 & \bf 0.000 & \bf 924 & & \bf 54--115 & \bf 5.5--9.0 & \bf 40--170  \\
\bf \stepcounter{BP}BP\arabic{BP} & \bf 5.4 & \bf 2.5 & \bf 1.0 & \bf 0.396 & \bf 445 & & \bf 5.4 & \bf 1.7 & \bf 1.40 & \bf 3.79 & \bf 0.000 & \bf 882 & & \bf 55--108 & \bf 5.5--8.5 & \bf 50--150  \\
\bf \stepcounter{BP}BP\arabic{BP} & \bf 5.4 & \bf 2.5 & \bf 1.1 & \bf 0.402 & \bf 399 & & \bf 5.4 & \bf 1.7 & \bf 1.35 & \bf 3.78 & \bf 0.004 & \bf 719 & & \bf 47--115 & \bf 5.5--9.0 & \bf 40--160 \\
\bf \stepcounter{BP}BP\arabic{BP} & \bf 5.4 & \bf 2.5 & \bf 1.2 & \bf 0.326 & \bf 436 & & \bf 5.4 & \bf 1.8 & \bf 1.36 & \bf 3.78 & \bf 0.022 & \bf 696 & & \bf 53--110 & \bf 5.5--9.0 & \bf 40--160 \\
\stepcounter{BP}BP\arabic{BP} & 5.4 & 2.5 & 1.3 & 0.293 & 461 & & 5.4 & 1.9 & 1.39 & 3.78 & 0.045 & 660 & & - & - & - \\
\stepcounter{BP}BP\arabic{BP} & 5.4 & 2.5 & 1.4 & 0.265 & 468 & & 5.4 & 1.9 & 1.43 & 3.79 & 0.067 & 618 & & - & - & - \\
\stepcounter{BP}BP\arabic{BP} & 5.4 & 3.0 & 0.6 & 0.699 & 5 & & 5.4 & 1.6 & 1.33 & 3.78 & 0.004 & 841 & & - & - & - \\
\bf \stepcounter{BP}BP\arabic{BP} & \bf 5.4 & \bf 3.0 & \bf 0.7 & \bf 0.612 & \bf  96 & & \bf 5.4 & \bf 1.7 & \bf 1.33 & \bf 3.77 & \bf 0.058 & \bf 679 & & \bf 51--108 & \bf 5.5--8.5  & \bf 40--170 \\
\bf \stepcounter{BP}BP\arabic{BP} & \bf 5.4 & \bf 3.0 & \bf 0.8 & \bf 0.548 & \bf 138 & & \bf 5.4 & \bf 1.8 & \bf 1.41 & \bf 3.79 & \bf 0.120 & \bf 485 & & \bf 52--106 & \bf 5.5--8.5 & \bf 40--160 \\
\stepcounter{BP}BP\arabic{BP} & 5.4 & 3.0 & 0.9 & 0.490 & 178 & & 5.4 & 1.9 & 1.48 & 3.80 & 0.158 & 417 & & - & - & - \\
\stepcounter{BP}BP\arabic{BP} & 5.4 & 3.3 & 0.6 & 0.700 & 1 & & 5.4 & 1.7 & 1.37 & 3.78 & 0.089 & 531 & & - & - & - \\
\stepcounter{BP}BP\arabic{BP} & 5.4 & 3.3 & 0.7 & 0.633 & 55 & & 5.4 & 1.8 & 1.47 & 3.80 & 0.163 & 400 & & - & - & - \\
\stepcounter{BP}BP\arabic{BP} & 5.4 & 3.3 & 0.8 & 0.566 & 98 & & 5.4 & 1.9 & 1.56 & 3.82 & 0.187 & 340 & & - & - & - \\
\stepcounter{BP}BP\arabic{BP} & 5.4 & 3.3 & 0.9 & 0.507 & 131 & & 5.4 & 2.0 & 1.63 & 3.84 & 0.190 & 309 & & - & - & - \\
%\stepcounter{BP}BP\arabic{BP} & 5.4 & 3.5 & 0.6 & 0.700 & 5 & & 5.4 & 1.8 & 1.45 & 3.80 & 0.152 & 418 & & - & - & - \\
%\stepcounter{BP}BP\arabic{BP} & 5.4 & 3.5 & 0.7 & 0.642 & 42 & & 5.4 & 1.9 & 1.56 & 3.82 & 0.189 & 332 & & - & - & - \\
%\stepcounter{BP}BP\arabic{BP} & 5.4 & 3.5 & 0.8 & 0.576 & 80 & & 5.4 & 2.0 & 1.65 & 3.84 & 0.195 & 290 & & - & - & - \\
%\stepcounter{BP}BP\arabic{BP} & 5.4 & 3.5 & 0.9 & 0.507 & 128 & & 5.4 & 2.3 & 1.83 & 3.88 & 0.175 & 302 & & - & - & - \\ 
\enddata
\end{deluxetable}
\clearpage

The shape of the evolutionary tracks in the H-R diagram is similar for
all considered sequences: the donor star follows its normal MS
evolution until the onset of MT (indicated by the filled circles) and
then evolves towards lower luminosities and lower effective
temperatures as it adjusts to its decreasing mass. The evolutionary
stage of the donor at the onset of MT is determined by the the orbital
period at the start of the RLO phase. For a given BH and donor mass,
longer initial periods yield more evolved donor stars and higher
initial MT rates. Since tracks with longer initial periods are also
associated with larger donor stars, they furthermore tend to be
shifted towards higher luminosities (cf.\ sequences 2 and 3). A
similar tendency occurs when increasing the donor mass for a fixed BH
mass and initial orbital period (cf.\ sequences 1, 3, and 4). The
requirement that the sequences must pass through the observationally
inferred position of GRO\,J1655-40 in the H-R diagram therefore gives
a rough {\em lower and upper limit on the initial period and donor
mass} at the onset of RLO.

For a given donor mass and a given initial period, the BH mass and the
amount of mass lost from the system during the MT process affect the
evolution in the H-R diagram through the rate at which the donor
reduces its mass. Higher BH masses and non-conservative MT yield lower
MT rates and therefore a slower decrease of the donor's luminosity and
effective temperature during the MT phase compared to that for lower
BH masses and conservative MT (cf. sequences 4, 5, and 6). These
effects are less severe though than changes in the initial donor mass
and orbital period.

The variations of the MT rate expressed in units of the critical rate
for transient behavior depend on the relative changes between these
two quantities. After the onset of the MT phase, the decrease of the
mass ratio and the increase of the orbital period cause the MT rate to
decrease and the critical rate to increase, so that overall the ratio
$\dot{M}/\dot{M}_{\rm crit}$ decreases with increasing orbital
period. This tendency continues until the donor reaches the end of
core-hydrogen burning and the associated expansion of the stellar
radius causes the MT rate to increase more rapidly than the critical
rate for transient behavior. The ratio $\dot{M}/\dot{M}_{\rm crit}$
therefore increases as well. During the subsequent evolution through
the Hertzsprung gap, no nuclear burning takes place and the MT rate
stays roughly constant. Since $\dot{M}_{\rm crit}$ still increases
slowly, the ratio $\dot{M}/\dot{M}_{\rm crit}$ slowly decreases.

Examination of Fig.\,\ref{tracks} indicates how we can effectively
identify {\em winner} sequences. All six sequences shown as an example
satisfy the GBO or BP constraints in the H-R diagram, but only two of
them (sequence 2 and 3) satisfy the BP mass constraints, and only one of
those satisfies the transient constraint (sequence 3).  The MT sequences
which are able to satisfy the constraints derived by GBO and BP are
highlighted in boldface in Tables~\ref{GBOseq} and~\ref{BPseq},
respectively.

\clearpage
\begin{figure*}
\resizebox{\hsize}{!}{\includegraphics{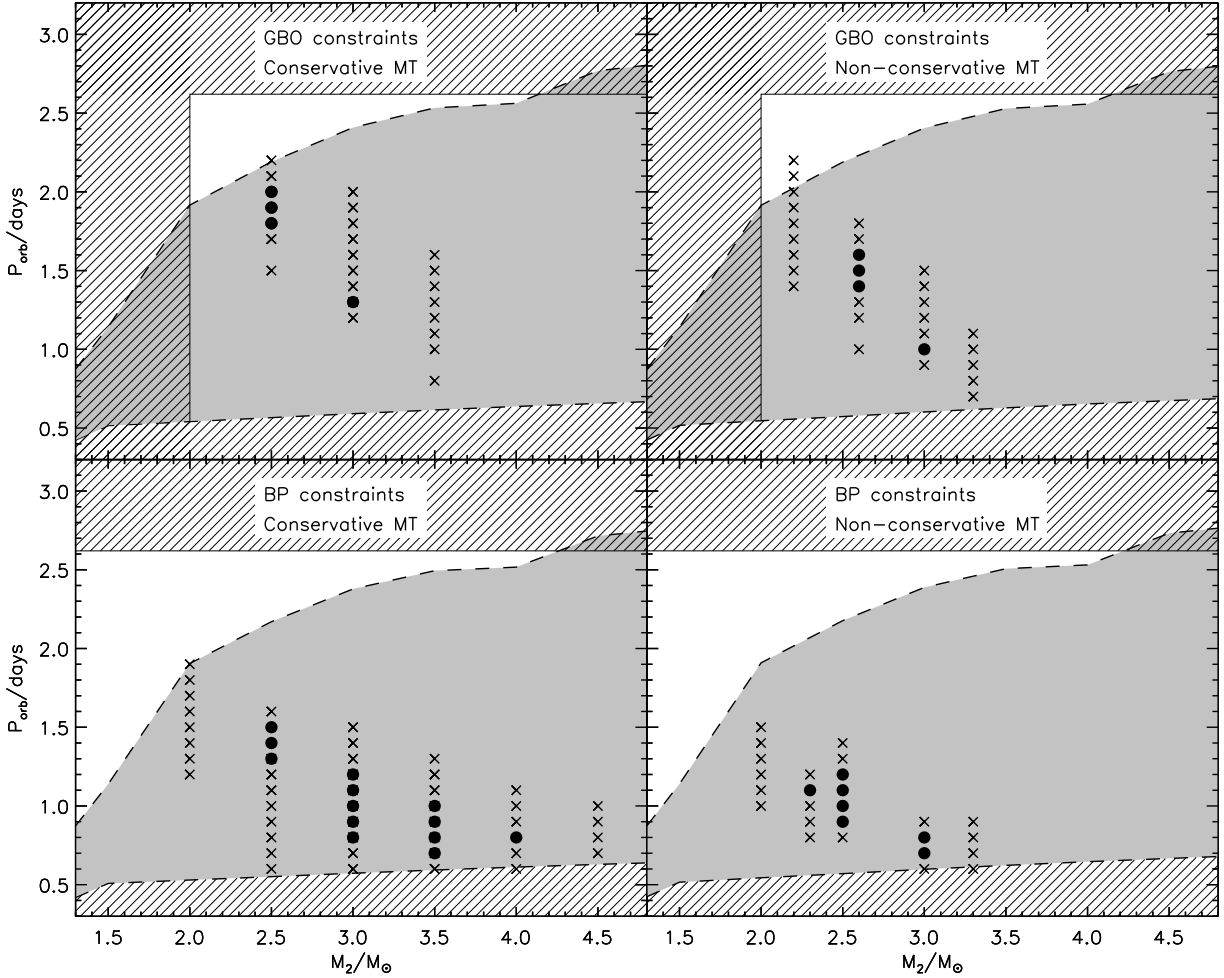}}
\caption{Initial donor masses and orbital periods of the evolutionary
   sequences at RLO onset. The grey shaded region delimited by the
   dashed lines indicates the region of the parameter space where the
   donor star is on the MS. Solid circles indicate MT sequences that
   are able to satisfy all the current observational constraints,
   while crosses indicate sequences that fail to do so. The hatched
   regions indicate regions of the parameter space that can be
   excluded without performing detailed MT calculations. For a given
   BH and donor mass, a lower limit on the initial period is given by
   the period at which the donor star fills its Roche lobe while it is
   still on the ZAMS. For MT sequences in which the orbital period
   increases as a result of MT, the current orbital period of
   2.62\,days furthermore provides an upper limit on the initial period. 
   In the case of GBO constraints, a lower limit on the mass is 
   given by the observational lower limit on the current donor mass of
   $2\,M_\odot$.}
\label{init}
\end{figure*}
\clearpage

In the case of GBO constraints and conservative MT, successful
sequences were found for initial BH masses ranging from $\simeq
5.5\,M_\odot$ to $6.0\,M_\odot$, initial donor masses ranging from
$\simeq 2.5\,M_\odot$ to $3.0\,M_\odot$, and initial periods ranging
from $\simeq 1.3$ to $2.0$\,days. The lower limit on the initial BH
mass arises from the requirement that, for conservative MT, lower mass
BHs need a more massive donor star in order to accrete enough matter
to reach the observed present-day BH mass. For orbital periods where
RLO occurs when the donor is still close to the ZAMS, the higher MT
rates associated with more massive donor stars then cause the donor
mass to decrease below the admissible range from 2.0 to2.8$\,M_\odot$
before the period reaches the present-day value of 2.62\,days. For
longer orbital periods, on the other hand, the donor star has a
somewhat more developed helium core at the onset of RLO, which causes
the star to be too luminous when the system reaches the observed
period of 2.62\,days. The upper limit on the initial BH mass is
determined by the requisite that as the initial BH mass gets closer to
the present value of $6.3 \pm 0.5\,M_\odot$, the donor star must start
RLO closer to its present state as well. This eventually requires MT
to start in the Hertzsprung gap which generally leads to MT rates that
are too high for the system to be transient (see, however, Kolb et
al. 1997 and Kolb 1998 for a discussion on soft X-ray transient in the
Hertzsprung gap). In the case of non-conservative MT, we kept the BH
mass fixed to the mean value of $6.3\,M_\odot$. By keeping the BH mass
constant we were able to to outline the initial donor masses at the
onset of RLO with a somewhat higher resolution of 0.3 to
$0.4\,M_\odot$ without having to calculate an unrealistically large
number of sequences. Successful sequences were found for $M_2 \simeq
2.6$--$3.0\,M_\odot$ and $P_{\rm orb} \simeq 1.0$--$1.6$\,days.  When
the donor star crosses the allowed region in the H-R diagram, all
successful MT sequences found for the GBO system parameters are
furthermore only marginally consistent with the currently observed
orbital period and donor mass. The MT rates at the period of
2.62\,days are always at least a factor of $\simeq 3$ smaller than
the critical MT rate for transient behavior.

In the case of BP constraints and conservative MT, successful
sequences were found for $M_{\rm BH} \simeq 3.5$--$4.5\,M_\odot$, $M_2
\simeq 2.5$--$4.0\,M_\odot$, and $P_{\rm orb} \simeq
0.7$--$1.5$\,days, while for non-conservative MT successful sequences
were found for $M_{\rm BH} \simeq 5.4\,M_\odot$, $M_2 \simeq
2.3\,M_\odot$--$3.0\,M_\odot$, and $P_{\rm orb} \simeq
0.7$--$1.2$\,days. The lower limit on the initial BH mass for
conservative MT is again related to the higher donor masses required
by initially less massive BHs to reach the present-day 
mass. The difference with the GBO sequences is, however, that the BH
masses at the start of RLO are now so low that this requirement
implies that MT initially starts on the thermal time scale of the
donor star. This leads to highly super-Eddington MT rates and
significant mass loss from the system. Consequently, initially even
more massive donor stars are required for the BH to accrete enough
to reach the observed present-day mass. These higher donor
masses increase the MT rates even further, rendering the system
persistent rather transient. The upper limit on the initial BH mass
arises for the same reason as for the GBO sequences. The MT rates
obtained when the donor star satisfies the observational constraints
in the H-R diagram and when the binary reaches the observed period of
2.62\,days span a wide range of values: for some sequences the MT
rates are more than an order of magnitude smaller than the critical
rate for transient behavior, while for others the system is only
marginally transient. In general, however, it turned out to be easier
to find MT sequences satisfying the BP constraints than sequences
satisfying the GBO constraints.

An overview of the initial donor star masses and the initial orbital
periods of the MT sequences listed in Tables~\ref{GBOseq}--\ref{BPseq}
is given in Fig.~\ref{init}. Successful sequences are indicated by
solid circles and unsuccessful ones by crosses. The figure clearly
illustrates the extent of the parameter space covered by our
calculations as well as the location of the successful sequences among
the unsuccessful ones. The general trend in the initial parameters of
the successful MT sequences is that lower initial donor masses require
longer initial orbital periods in order to be successful. This is in
agreement with the systematic behavior of the MT sequences illustrated
in Fig.~\ref{tracks}. We also note that MT sequences were calculated
for BH masses outside the successful ranges quoted above, but for
brevity these have been omitted from Tables~\ref{GBOseq}--\ref{BPseq}
and Fig.~\ref{init}.

Fig.~\ref{init} and Tables~\ref{GBOseq}--\ref{BPseq} also show there
is no single {\em unique} MT sequence that satisfies all the
observational constraints. Instead, a series of neighboring successful
sequences is found for both GBO and BP constraints and for both
conservative and non-conservative MT. In addition, the number of
successful sequences found is limited only by our finite exploration
of the available parameter space and the limitation of the
calculations to the two extreme cases of fully conservative and fully
non-conservative MT. Adopting a higher resolution grid of initial
component masses and orbital periods would increase the number of
successful sequences even further as would the consideration of MT
sequences for which the degree of mass conservation is in between the
two extreme cases of fully conservative or non-conservative
MT. Fortunately, as we will see in Section~\ref{prog}, MT sequences
with closely spaced initial parameters yield only small variations in
the properties of the pre-SN binary progenitor and the kick that may
have been imparted to the BH at birth. The vital point in the
calculation of the presented MT sequences is therefore that we managed
to outline the initial parameters of the successful sequences within a
pre-determined accuracy rather than finding all possible MT sequences
able to satisfy the present-day observational constraints for
GRO\,J1655-40.

\section{KINEMATIC HISTORY IN THE GALAXY}
\label{galmotion}

Given the present position and velocity of GRO\,J1655-40 in the
Galaxy, we derive the post-SN peculiar velocity of the binary's mass
center by tracing its orbit in the Galaxy back to the birth time of
the BH. A specific birth time is derived for {\em each} successful MT
sequence by setting it equal to the {\em current} age of the
donor star. The latter is a good estimate of the BH birth time within
an uncertainty equal to the lifetime of the massive progenitor of the
BH, which is typically less than 10\,Myr. This uncertainty is always
much smaller than the actual time expired since the birth of the BH.

We describe the motion of the system with respect to a right-handed
Cartesian frame of reference $OXYZ$ whose origin coincides with the
Galactic center and whose $XY$-plane coincides with the mid-plane of
the Galactic disk. The direction from the projection of the Sun's
position onto the Galactic plane to the Galactic center is taken as
the positive direction of the $X$-axis, the direction from the Sun to
the Northern Galactic pole as the positive direction of the $Z$-axis,
and the direction of the Galactic rotational velocity at the position
of the Sun as the positive direction of the $Y$-axis. In terms of
these coordinates, the Sun is located at $\left( X_\odot, Y_\odot,
Z_\odot\right)$ with $X_\odot = -8 \pm 0.5$\,kpc, $Y_\odot = 0$, and
$Z_\odot = 30 \pm 25$\,pc (Reid 1993; Humphreys \& Larsen 1995, and
references therein). GRO\,J1655-40 is currently located at a distance 
from the Sun of $3.2 \pm 0.2$\,kpc (Hjellming \& Rupen 1995), a 
Galactic longitude $l = 345.0^\circ$, and a Galactic latitude 
$b = +2.2^\circ$ (Tingay et al. 1995); corresponding to $X = -4.9 \pm 
0.5$\,kpc, $Y = -0.83 \pm 0.05$\,kpc, and $Z = 0.15 \pm 0.03$\,kpc.
The velocity components of GRO\,J1655-40 with respect to
the $X$-, $Y$-, and $Z$-axes are $U = -121 \pm 18\,{\rm km\,s^{-1}}$,
$V = -33 \pm 8\,{\rm km\,s^{-1}}$, and $W = 3 \pm 8\,{\rm km\,s^{-1}}$
(Mirabel et al. 2002).

To model the Galaxy, we adopt the Galactic potential of Carlberg \&
Innanen (1987) with updated model parameters of Kuijken \& Gilmore
(1989). For each successful MT sequence (marked in boldface in
Tables~\ref{GBOseq} and~\ref{BPseq}), the equations governing the
system's motion in the Galaxy are integrated backward in time up to
the time corresponding to the current age of the donor star, as given
by the evolutionary sequence. 

\clearpage
\begin{figure*}
\resizebox{\hsize}{!}{\includegraphics{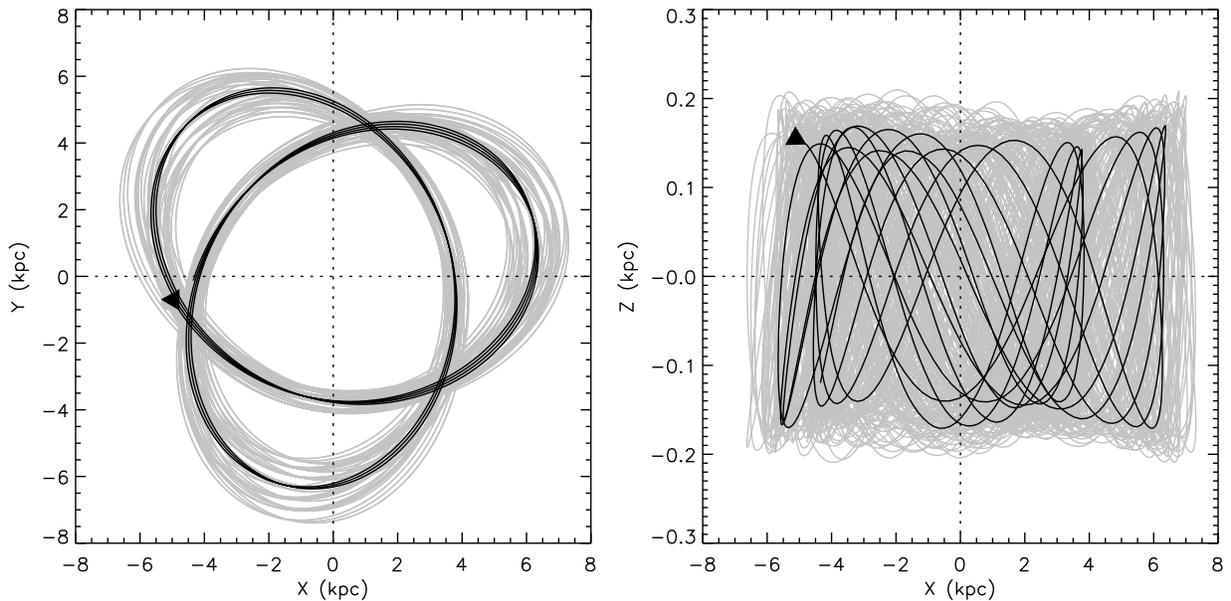}}
\caption{Orbit of GRO\,J1655-40 in the Galaxy up to 1\,Gr in the
    past. In the left-hand panel, the orbit is shown projected on the
    $XY$-plane, while in the right-hand panel it is projected on the
    $XZ$-plane (see text for a definition of the coordinate axes). The
    black line indicates the orbit for a present-day distance of
    3.2\,kpc and velocity components $U = -121\,{\rm km\,s^{-1}}$, $V
    = -33\,{\rm km\,s^{-1}}$, and $W = 3\,{\rm km\,s^{-1}}$. The grey
    lines indicate the uncertainties in the orbit associated with the
    error bars in the distance and the velocity components. The arrow
    indicates the present-day position of GRO\,J1655-40 and the
    direction of the motion backward in time.}
\label{orbit}
\end{figure*}
\clearpage

In Fig.~\ref{orbit}, we show the orbit of GRO\,J1655-40 projected on
the $XY$- and $XZ$-planes up to 1\,Gyr in the past, which encompasses
the entire range of donor ages from $\simeq 335$ to $\simeq 998$\,Myr
given by the successful MT sequences. The black line represents the
orbit obtained for the mean values of the distance and velocity
components, while the grey lines indicate the deviations from this
orbit obtained from considering all possible combinations of the
extreme values of the distance and velocity component measurements
(i.e. the endpoints of the error bars). In agreement with the findings
of Mirabel et al. (2002), the system always remains within 3 to 7\,kpc
from the Galactic center and within 200\,pc from the Galactic plane.

\clearpage
\begin{figure*}
\resizebox{\hsize}{!}{\includegraphics{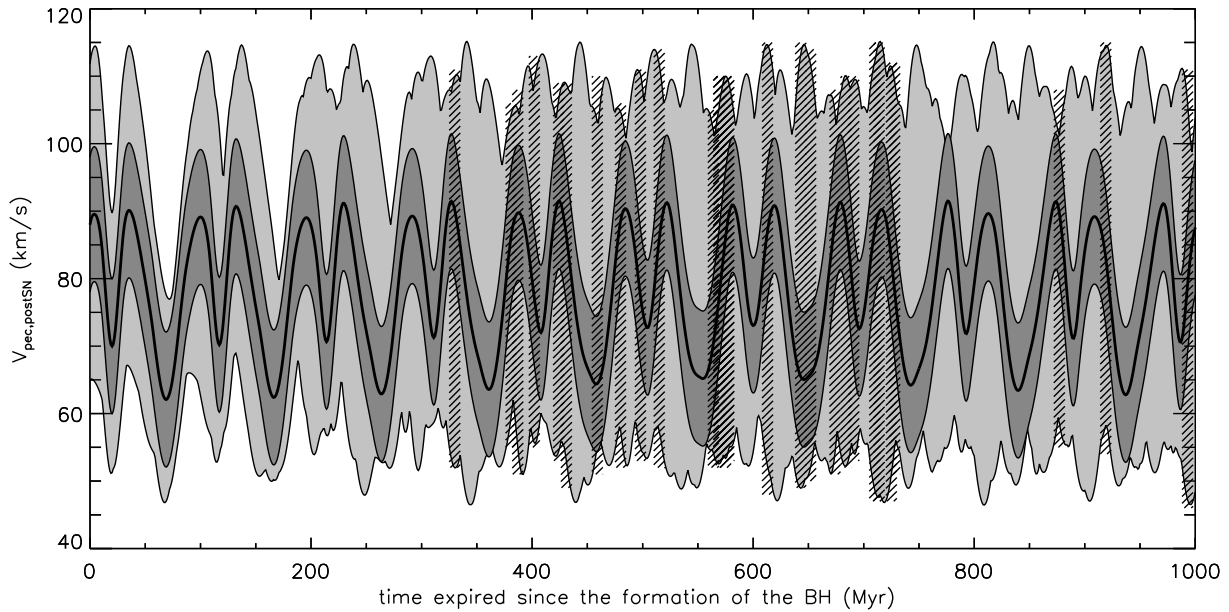}}
\caption{Post-SN peculiar velocity of GRO\,J1655-40 as a function of
  the time expired since the formation of the BH. The black solid line
  represents the post-SN peculiar velocity for the mean values of the
  binary's current distance and 3D velocity, and the light grey area
  indicates the uncertainties resulting from the error bars in the
  distance and the velocity components. The dark grey area indicates
  the $\pm 10\,{\rm km\,s^{-1}}$ errors bars on $V_{\rm
  pec,postSN}$. Possible birth times of the BH given by the successful
  MT sequences listed in Tables~\ref{GBOseq} and~\ref{BPseq} are
  indicated by the hatched vertical bars.}
\label{vcm}
\end{figure*}
\clearpage

The post-SN peculiar velocity $V_{\rm pec,postSN}$ of the binary is
obtained by subtracting the local Galactic rotational velocity from
the total systemic velocity at the birth site of the BH. The
variations of $V_{\rm pec,postSN}$ as a function of the time expired
since the formation of the BH are displayed in Fig.~\ref{vcm}. The
hatched vertical bars indicate the BH birth times determined by the
age of the donor star for each of the successful MT sequences listed
in Tables~\ref{GBOseq}--\ref{BPseq}. The width of the bars corresponds
to an uncertainty in the birth times of 10\,Myr. When accounting for
the uncertainties in the distance and the velocity components
(cf. Fig.~\ref{orbit}), the post-SN peculiar velocity can span a wide
range of values ranging from $\simeq 45$ to $\simeq 115\,{\rm
km\,s^{-1}}$. This wide range is a result of our determination of the
possible uncertainties in the Galactic orbit which likely
overestimates the true error bars on $V_{\rm
pec,postSN}$. Nevertheless, in order to be conservative, we consider
the full range of possible post-SN peculiar velocities given by the
grey shaded area in Fig.~\ref{vcm}.

\section{ORBITAL EVOLUTION DUE TO TIDES AND GENERAL RELATIVITY }
\label{tidaleq}

After the formation of the BH, the orbital parameters of the binary
are subjected to secular changes due to the perturbation of the
external gravitational field from spherical symmetry by the MS star's
tidal distortion and due to the loss of orbital angular momentum via
gravitational radiation. Since the tidal interactions depend on both
the orbital and rotational properties of the MS star, the star's
rotational angular velocity right after the SN explosion that formed
the BH enters the problem as an additional unknown quantity. We here
assume that the the rotational angular velocity of the MS star is
unaffected by the SN explosion and that just before the explosion the
star's rotation rate was synchronized with the orbital motion. The
post-SN rotational angular velocity is then given by $\Omega_{\rm
postSN} = 2\,\pi/P_{\rm orb,preSN}$, where $P_{\rm orb,preSN}$ is the
pre-SN orbital period.  As a test, we also considered the more extreme
cases $\Omega_{\rm postSN} = 0$ and $\Omega_{\rm postSN} = 10 \times
2\,\pi/P_{\rm orb,preSN}$, but these were found to not affect
significantly the outcome of the orbital evolution between the
formation of the BH and the start of the RLO phase.

We adopt the system of equations governing the tidal evolution of the
semi-major axis $A$, the orbital eccentricity $e$, and the
MS star's rotational angular velocity $\Omega$ from Hut (1981):
 \begin{eqnarray}
\lefteqn{\left( {{dA} \over {dt}} \right)_{\rm tides} = 
   -6\, {k_2 \over T}\, 
   {M_{\rm BH} \over M_2}\, {{M_{\rm BH}+M_2} \over M_2} 
   \left( {R_2 \over A} \right)^8}   \nonumber \\
 & & \times {A \over {\left( 1 - e^2 \right)^{15/2}}}
   \left[ f_1 \left( e^2 \right) - \left( 1 - e^2 
   \right)^{3/2} f_2 \left( e^2 \right) {\Omega \over n}
   \right],
\end{eqnarray}
\begin{eqnarray}
\lefteqn{\left( {{de} \over {dt}} \right)_{\rm tides} = 
   -27\, {k_2 \over T}\, 
   {M_{\rm BH} \over M_2}\, {{M_{\rm BH}+M_2} \over M_2} 
   \left( {R_2 \over A} \right)^8}   \nonumber \\
 & & \times {e \over {\left( 1 - e^2 \right)^{13/2}}}
   \left[ f_3 \left( e^2 \right) - {{11} \over {18}} 
   \left( 1 - e^2 \right)^{3/2} f_4 \left( e^2 \right) 
   {\Omega \over n} \right],
\end{eqnarray}
\begin{eqnarray}
\lefteqn{\left( {{d\Omega} \over {dt}} \right)_{\rm tides} = 
   3\, {k_2 \over T} 
   \left( {M_{\rm BH} \over M_2} \right)^2
   {{M_2\, R_2^2} \over I_2}
   \left( {R_2 \over A} \right)^6}  \nonumber \\
 & & \times {n \over {\left( 1 - e^2 \right)^6}}
   \left[ f_2 \left( e^2 \right) -  
   \left( 1 - e^2 \right)^{3/2} f_5 \left( e^2 \right) 
   {\Omega \over n} \right],
\end{eqnarray}
where
\begin{equation}
f_1 \left( e^2 \right) = 1 + {{31} \over 2}\, e^2
   + {{255} \over 8}\, e^4 + {{185} \over {16}}\, e^6
   + {{25} \over {64}}\, e^8,
\end{equation}
\begin{equation}
f_2 \left( e^2 \right) = 1 + {{15} \over 2}\, e^2
   + {{45} \over 8}\, e^4 + {5 \over {16}}\, e^6,
\end{equation}
\begin{equation}
f_3 \left( e^2 \right) = 1 + {{15} \over 4}\, e^2
   + {{15} \over 8}\, e^4 + {5 \over {64}}\, e^6,
\end{equation}
\begin{equation}
f_4 \left( e^2 \right) = 1 + {3 \over 2}\, e^2
   + {1 \over 8}\, e^4,
\end{equation}
\begin{equation}
f_5 \left( e^2 \right) = 1 + 3\, e^2
   + {3 \over 8}\, e^4
\end{equation}
(see also Zahn 1977, 1978). In these equations, $M_{\rm BH}$ is the
mass of the BH; $M_2$, $R_2$, $k_2$, and $I_2$ are the mass, radius,
apsidal-motion constant, and moment of inertia of the companion star,
$n=2\,\pi/P_{\rm orb}$ is the mean orbital angular velocity, and $T$
is a characteristic time scale for the orbital evolution due to tides.

For stars with radiative envelopes, as we are dealing with here, the
dominant mechanism contributing to the dissipative tidal forces is
radiative damping of dynamical tides (Zahn 1975, 1977). The factor
$k_2/T$ can then be approximated as
\begin{eqnarray}
\left( {k_2 \over T} \right)_{\rm rad} & & = 
  1.9782 \times 10^4\, f_{\rm cal}
  \left( R_2 \over R_\odot\right)^2
  \left( R_\odot \over A \right) 
  \nonumber \\
 & & \times \left( M_2 \over M_\odot \right)
  \left( {{M_{\rm BH}+M_2} \over M_2} \right)^{5/6} 
  E_2 \,\, {\rm yr^{-1}}  \hspace{1.0cm}
  \label{kT}
\end{eqnarray}
with
\begin{equation}
E_2 = 1.592 \times 10^{-9} \left( M_2 \over M_\odot \right)^{2.84}
   \label{E2}
\end{equation}
(Hurley, Tout, \& Pols 2002).

Eqs. (\ref{kT}) and (\ref{E2}) are strictly speaking only valid for
dynamic tides with low forcing frequencies that are far away from any
of the eigenfrequencies of the star's free modes of oscillation. Witte
\& Savonije (1999) have shown that in close binaries with eccentric
orbits the system can easily become locked in a resonance between a
dynamic tide and a free oscillation mode and that such resonance
lockings can greatly accelerate the orbital evolution. In the
expression for $k_2/T$, we therefore incorporated the calibration
factor $f_{\rm cal}$ to account for some of the still existing
uncertainties in the strength of tidal dissipation in stars containing
radiative envelopes. In our calculations, we adopt $f_{\rm cal}=10$
which has been shown to be appropriate for consistency with both the
observations of circularization periods in open clusters and with
orbital decay rates measured in high-mass X-ray binaries (for more
details see Belczynski et al.\ 2004).

To follow the secular changes in the orbital parameters associated
with the emission of gravitational waves, we adopt the system of
equations given by Junker \& Sch\"{a}fer (1992):
\begin{eqnarray}
\left( {{dA} \over {dt}} \right)_{\rm GR} & & = 
  - {{2\,c} \over {15}}\, 
  {\nu \over {\left( 1-e^2 \right)^{9/2}}}
  \left[ {{G \left( M_{\rm BH} + M_2 \right)} \over 
  {A\, c^2}} \right]^3  \nonumber \\
 & & \times \left( 96 + 292\,e^2 + 37\,e^4 
  \right) \left( 1 - e^2 \right)
  \label{dagr}
\end{eqnarray}
\begin{eqnarray}
\left( {{de} \over {dt}} \right)_{\rm GR} & & = 
  - {1 \over {15}}\, 
  {{\nu\, c^3} \over {G \left( M_{\rm BH} + M_2 \right)}}
  \left[ {{G \left( M_{\rm BH} + M_2 \right)} \over 
  {A\, c^2}} \right]^4  \nonumber \\
 & & \times {{e \left( 304 + 121\,e^2 \right) 
  \left( 1 - e^2 \right)} \over {\left( 1-e^2 \right)^{7/2}}}. 
  \label{degr}
\end{eqnarray}
Here, $G$ is the Newtonian gravitational constant, $c$ is the speed
of light, and $\nu = M_{\rm BH}\,M_2/(M_{\rm BH} + M_2)^2$. For
brevity, we restricted Eqs. (\ref{dagr}) and (\ref{degr}) to the
lowest-order terms in the small parameter $G(M_{\rm BH} + M_2)/(A\,
c^2)$. In the numerical calculations however, we adopt the full 3.5
post-Newtonian order equations derived by Junker \& Sch\"{a}fer
(1992).

\begin{figure*}
\resizebox{\hsize}{!}{\includegraphics{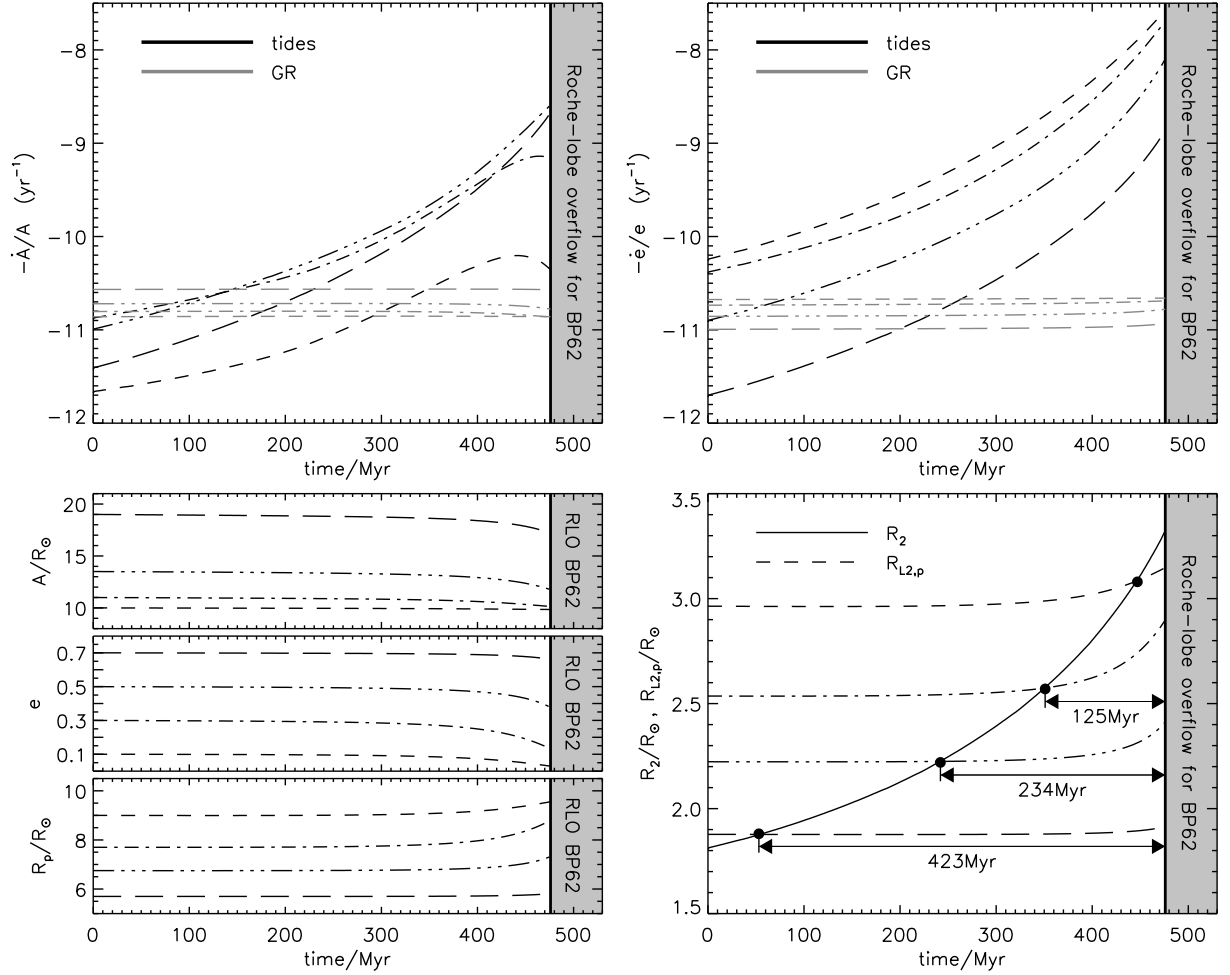}}
\caption{Secular evolution of a binary consisting of a $4.5\,M_\odot$
BH and a $2.5\,M_\odot$ ZAMS star, for four different combinations of
initial orbital separations and eccentricities (see the bottom left
panels to identify the combinations and the associated line
types). The top panels show the relative rates of secular change of
the semi-major axis $A$ and the orbital eccentricity $e$ as a result
of the MS star's tidal distortion and gravitational radiation
separately. The bottom-left panels show the evolution of the
semi-major axis, orbital eccentricity, and periastron distance
$R_p=A(1-e)$ resulting from the combined effects of tides and
gravitational radiation, while the bottom-right panel shows the
associated evolution of the MS star's Roche-lobe radius at the
periastron of its relative orbit. For comparison, the evolution of the
star's radius is also shown in the bottom-right panel.  The thick
vertical lines followed by the grey-shaded areas mark the beginning of
the RLO phase for the successful MT sequence
BP\ref{example1}.}
\label{orbev}
\end{figure*}

The orbital evolution due to tides and gravitational radiation is
illustrated in Fig.~\ref{orbev} for the example case of a binary
consisting of a $4.5\,M_\odot$ BH and a $2.5\,M_\odot$ zero-age MS
star, and for four different combinations of initial orbital separations
and eccentricities. The four combinations are chosen such that after
approximately $\simeq 480$\,Myr of orbital evolution the total angular
momentum of the binary is approximately equal to the total angular
momentum of the binary represented by MT sequence BP\ref{example1}
(see Table~\ref{BPseq}) at the onset of the MT phase\footnote{Note
that 480\,Myr corresponds to the age of the donor star at the onset of
MT in sequence BP\ref{example1}. The choice to match the orbital
angular momentum at the end of the orbital evolution calculations to
the orbital angular momentum of the MT sequence at the onset of RLO is
equivalent to assuming that tides circularize the binary
instantaneously at $t \simeq 480$\,Myr.}.

\clearpage
\clearpage

The two upper panels of Fig.~\ref{orbev} show the relative rates of
secular change of the semi-major axis $A$ and the orbital eccentricity
$e$ (for tides and GR separately) as functions of the time expired
since the formation of the BH. For all four combinations of initial
orbital separations and eccentricities, the GR rates remain almost
constant throughout the considered time interval, while the rates due
to tides increase significantly as the star evolves on the MS. This
behavior reflects the independence of gravitational radiation on the
size and structure of the star and the strong dependence of tides on
the ratio of the stellar radius to the orbital semi-major axis. The
plots also illustrate that, for binaries consisting of a compact
object and a young early-type secondary, gravitational radiation and
tides, at some points in time, may affect the orbital evolution to
comparable degrees.

The actual evolution of the semi-major axis $A$, the orbital
eccentricity $e$, and the periastron distance $R_p=A(1-e)$ due to the
combined effect of tides and gravitational radiation is displayed in
the lower-left panels of Fig.\,\ref{orbev}. As expected from the low
rates of secular change displayed in the upper two panels, the orbital
parameters remain almost constant during the first 300-400\,Myr of
evolution and start changing noticeably when the star gets closer to
the end of the MS and tidal effects become dominant.

Finally, in the lower-right panel of Fig.~\ref{orbev} we compare the
evolution of the stellar radius with the evolution of the Roche-lobe
radius at the periastron of the MS star's relative orbit for each of
the four considered combinations of initial orbital separations and
eccentricities. The Roche-lobe radii at periastron are determined by
\begin{equation}
{R_{{\rm Lp},2} \over R_p} = {{0.49\, q^{2/3}} \over {0.6\, q^{2/3}
  + \ln \left( 1 + q^{1/3} \right)}},  \label{RLp}
\end{equation}
where $q = M_2/M_{\rm BH}$ (Eggleton 1983). It follows that although
all four pairs of initial orbital separations and eccentricities lead
to orbital configurations that are consistent with the successful MT
sequence BP\ref{example1} at the onset of RLO, only one pair
($A=10\,R_\odot$, $e=0.1$) leads to RLO at periastron at a time that
is sufficiently close to the age of the donor star required to make
the sequence successful in the first place. The other three pairs all
yield RLO at periastron at times that are $\simeq 125-420$\,Myr too
early. They would therefore only yield viable progenitors for
GRO\,J1655-40 if RLO at periastron is assumed to conserve the binary's
total angular momentum and if the total amount of mass transferred
from the donor star is small enough to not significantly change its
properties so that it would follow a different evolutionary path in
the H-R diagram. If either of these conditions is violated, the
considered pair of initial orbital separation and eccentricity is
incompatible with the starting point of the successful RLO sequence
under consideration.

\section{ORBITAL DYNAMICS AT CORE COLLAPSE}

At the time of the SN explosion, the mass lost from the system and
possibly the kick imparted to the BH during the explosion change the
binary's orbital parameters. The pre- and post-SN component masses,
orbital semi-major axis, and orbital eccentricity are related by the
conservation laws of orbital energy and angular momentum, which in
turn depend on the magnitude and the direction of the kick velocity
that may be imparted to the BH. We describe these relations and the
associated progenitor and kick velocity constraints in more detail in
Sections~\ref{symmsn} and~\ref{asymmsn}. In these sections, we adopt
the same notations for the orbital elements as before, but add the
subscripts "preSN" and "postSN" to distinguish between their values
before and after the SN explosion that formed the BH.

We first note though that there are two additional constraints that
need to be imposed on the BH progenitor besides those arising from the
orbital dynamics of symmetric and asymmetric SN explosions. One is
that, for each successful MT sequence, the progenitor mass must of
course be more massive than the initial BH mass at the onset of
RLO. The other is that the BH progenitor must fit within its pre-SN
Roche lobe. Strictly speaking this is not a necessary condition (Dewi
\& van den Heuvel 2004; Willems \& Kalogera 2004; and Willems et
al. 2004, for example, showed that the progenitor of PSR~J0737-3039B
was filling its Roche lobe at the time of its collapse). Nevertheless,
we impose this condition to avoid complications arising from
accretion-induced changes in the stellar structure of the MS companion
that will become the XRB donor star during the X-ray phase. We have
tested this assumption by omitting the second constraint from our
calculations and found no appreciable change in the derived progenitor
and kick velocity constraints.

\subsection{Symmetric SN explosions}
\label{symmsn}

The orbital dynamics of symmetric SN explosions have been discussed
extensively in the past (e.g., Blaauw 1961; Boersma 1961; Hills 1983;
Fryer \& Kalogera 1997). We recall that the changes in the pre-SN
orbital parameters are entirely determined by the amount of mass lost
from the system during the explosion, yielding a post-SN semi-major
axis and orbital eccentricity given by
\begin{equation}
  {A_{\rm postSN} \over A_{\rm preSN}} = 
  {{M_{\rm BH}+M_2} \over {M_2+2\,M_{\rm BH}-M_{\rm He}}},
  \label{sym1}
\end{equation}
\begin{equation}
  e_{\rm postSN} = {{M_{\rm He} - M_{\rm BH}} \over {M_{\rm BH} + M_2}},
  \label{sym2}
\end{equation}
where $M_{\rm He}$ is the mass of the BH's helium star progenitor.

The mass loss from the system also imparts a kick to the binary's
center of mass with a magnitude equal to 
\begin{equation}
  V_{\rm pec,postSN} = {{\left( M_{\rm He} - M_{\rm BH} 
  \right) M_2} \over {\left( M_{\rm BH} + M_2 \right)
  \left( M_{\rm He} + M_2 \right)}}\, V_{\rm He,preSN},
  \label{sym3}  
\end{equation}
where 
\begin{equation}
V_{\rm He,preSN} = \left[ {{G \left( M_{\rm He} + M_2 \right)}  
  \over A_{\rm preSN}} \right]^{1/2}  \label{sym4}
\end{equation}
is the relative orbital velocity
of the helium star just before its instantaneous SN explosion. 

\subsection{Asymmetric SN explosions}
\label{asymmsn}

For asymmetric SN explosions, the newly formed BH receives a
kick at birth, the magnitude and direction of which can either
counteract or reinforce the effect of the mass lost from the
system. The pre- and post-SN binary parameters are related by the
equations  
\begin{eqnarray}
\lefteqn{V_{\rm k}^2 + V_{\rm He,preSN}^2 
 + 2\, V_{\rm k}\, V_{\rm He,preSN}\, \cos
 \theta} \nonumber \\
 & & = G \left( M_{\rm BH} + M_2 \right) \left( {2 \over A_{\rm
 preSN}} - {1 \over A_{\rm postSN}} \right),  \label{eq1}
\end{eqnarray}
\begin{eqnarray}
\lefteqn{A_{\rm preSN}^2 \left[ V_{\rm k}^2\, \sin^2 \theta \cos^2
 \phi \right. + \left. \left( V_{\rm k}\, \cos \theta 
 + V_{\rm He,preSN} \right)^2 \right]} \nonumber \\
 & = & G \left( M_{\rm BH} + M_2 \right) A_{\rm postSN} 
 \left( 1 - e_{\rm postSN}^2 \right) \hspace{1.2cm} \label{eq2}
\end{eqnarray}
(e.g., Hills 1983; Brandt \& Podsiadlowski 1995; Kalogera 1996; Fryer
\& Kalogera 1997; Kalogera \& Lorimer 2000). Here, $V_k$ is the
magnitude of the kick velocity, $\theta$ is the polar angle between
the kick velocity and the relative orbital velocity of the helium star
just before the SN explosion, and $\phi$ is the corresponding
azimuthal angle defined so that $\phi=0$ represents a plane
perpendicular to the line connecting the centers of mass of the binary
components (see Fig.~1 in Kalogera 2000 for a graphical representation).

The requirements that the binary must remain bound after the SN
explosion and that the direction of the kick must be real ($0 \le
\sin^2 \theta \le 1$ and $0 \le \cos^2 \phi \le 1$) impose constraints
on the pre- and post-SN parameters and on the magnitude of the kick
velocity imparted to the BH at birth. In particular, the ratio of the
pre- to post-SN semi-major axes must satisfy the inequalities
\begin{equation}
1-e_{\rm postSN} \le {A_{\rm preSN}
  \over A_{\rm postSN}} \le 1-e_{\rm postSN}, 
\end{equation}
\begin{equation}
{A_{\rm preSN} \over A_{\rm postSN}} < 2 - {{M_{\rm He}+M_2} \over
  {M_{\rm BH}+M_2}} \left( {V_k \over V_{\rm He,preSN}} - 1 \right)^2, 
\end{equation}
\begin{equation}
{A_{\rm preSN} \over A_{\rm postSN}} > 2 - {{M_{\rm He}+M_2} \over
  {M_{\rm BH}+M_2}} \left( {V_k \over V_{\rm He,preSN}} + 1 \right)^2.
\end{equation}
The first of these inequalities expresses the condition that the
post-SN orbit must pass through the position of the two stars at the
time of the SN explosion (Flannery \& van den Heuvel 1975). The last
two inequalities correspond to lower and upper limits on the amount of
orbital contraction or expansion that can take place for a given
amount of mass loss and a given magnitude of the kick velocity
(see, e.g., Kalogera \& Lorimer 2000).  

The magnitude of the kick velocity imparted to the BH at birth is
restricted to the range determined by (Brandt \& Podsiadlowski 1995;
Kalogera \& Lorimer 2000)
\begin{equation}
{V_k \over V_{\rm He,preSN}} < 1+ \left( 2\,
  {{M_{\rm BH}+M_2} \over {M_{\rm He}+M_2}} \right)^{1/2},
\end{equation}
\begin{equation}
{V_k \over V_{\rm He,preSN}} > 1 - \left( 2\,
  {{M_{\rm BH}+M_2} \over {M_{\rm He}+M_2}} \right)^{1/2}.
\end{equation}
The first inequality expresses the requirement that the binary must
remain bound after the SN explosion, while the second inequality
yields the minimum kick velocity required to keep the system bound if
more than half of the total system mass is lost in the explosion.

Last, an upper limit on the mass of the BH progenitor can be derived
from the condition that the azimuthal direction of the kick is real, 
i.e., $\cos^2 \phi \ge 0$ (Fryer \& Kalogera 1997):
\begin{eqnarray}
  M_{\rm He} & & \le -M_2 + k^2 \left( M_2 + M_{\rm BH} 
  \right) {A_{\rm preSN} \over A_{\rm postSN}}  \nonumber \\
  & & \! \times \Bigg\{ 2 \left( {A_{\rm postSN} 
  \over A_{\rm preSN}} \right)^2 
  \left( 1 - e_{\rm postSN}^2 \right) - k  \nonumber \\
  & & - 2\, {A_{\rm postSN} \over A_{\rm preSN}}
  \left( 1 - e^2 \right)^{1/2} \\
  & & \times \left[ 
  \left( {A_{\rm postSN} \over A_{\rm preSN}} \right)^2 
  \left( 1 - e_{\rm postSN}^2 \right) - k \right]^{1/2}
  \Bigg\}^{-1},  \nonumber 
\end{eqnarray}
where
\begin{equation}
  k = 2\, {A_{\rm postSN} \over A_{\rm preSN}} - 
  \left[ {{V_{\rm k}^2\, A_{\rm postSN}} \over 
  {G \left( M_2 + M_{\rm BH} \right)}} + 1 \right].
\end{equation}

From Eqs.~(3) and (34) in Kalogera (1996), it also follows that
the mass loss and kick experienced during the SN explosion yield a
post-SN peculiar velocity for the binary's center of mass determined
by 
\begin{equation}
  V_{\rm pec,postSN}^2 = \nu_1\, V_{\rm He,preSN}^2 + \nu_2\, V_{\rm
  BH,postSN}^2 + \nu_3\, V_k^2,
  \label{vpostSN}
\end{equation}
where 
\begin{equation}
V_{\rm BH,postSN}^2 = G \left( M_{\rm BH} + M_2 \right)
  \left( {2 \over A_{\rm preSN}} - {1 \over A_{\rm postSN}} \right)
\end{equation}
is the square of the relative orbital velocity of the BH {\it 
immediately after} the SN explosion. The coefficients 
$\nu_1$, $\nu_2$, and $\nu_3$ are defined as
\begin{equation}
  \nu_1 = {{M_{\rm He}\, M_2 \left( M_{\rm He} - M_{\rm BH} \right)}
  \over {\left( M_{\rm BH} + M_2 \right)  
  \left( M_{\rm He} + M_2 \right)^2}},
\end{equation}
\begin{equation}
  \nu_2 = - {{M_{\rm BH}\, M_2 \left( M_{\rm He} - M_{\rm BH} \right)}
  \over {\left( M_{\rm BH} + M_2 \right)^2 \left( M_{\rm He} + M_2
  \right)}}, 
\end{equation}
\begin{equation}
  \nu_3 = {{M_{\rm BH}\, M_{\rm He}} \over {\left( M_{\rm BH} + M_2
  \right) \left( M_{\rm He} + M_2 \right)}}.  \label{nu3}
\end{equation}

For each successful MT sequence, we use the post-SN orbital parameters
derived by following the Galactic motion and the orbital evolution due
to tides and gravitation radiation backward in time
(see Sections~\ref{galmotion} and~\ref{tidaleq}) and explore the
parameter space made up by $A_{\rm preSN}$, $M_{\rm He}$, $V_k$,
$\theta$, and $\phi$ to look for pre-SN binary configurations and kick
parameters satisfying Eqs.~(\ref{eq1})--(\ref{nu3}), or, in the case
of $V_k=0$, Eqs.~(\ref{sym1})--(\ref{sym4}). As we will see in the
next section, this allows us to identify well defined regions of
progenitor and kick properties that are fully consistent with the
complete set of observational constraints described in
Section~\ref{gro1655}.  

\section{PROGENITOR CONSTRAINTS} 
\label{prog}

The elements presented in the previous sections can now be combined to
establish a complete picture of the evolutionary history of
GRO\,J1655-40, the pre- and post-SN binary properties, and the
dynamics involved in the core-collapse event that formed the BH. For
{\em each} of the MT sequences that reproduce all of the system's
currently observed properties, we first trace both the motion of the
system in the Galaxy and the orbital evolution due to tides and
gravitational radiation back in time to the formation of the BH. The
BH formation time is different for each MT sequence and is estimated
by the age of the donor star. This procedure gives
us the binary's peculiar velocity and orbital parameters right after
the BH formation. We then use the conservation laws of orbital
energy and orbital angular momentum together with all the associated
orbital dynamics constraints at core collapse to link the pre-SN
binary properties and the possible BH kick to the post-SN binary
properties. In the following paragraphs, we first, as an example,
discuss the progenitor constraints derived following this procedure
for a single successful MT sequence, and next summarize the
conclusions from all successful MT sequences.

\clearpage
\begin{figure*}
\resizebox{\hsize}{!}{\includegraphics{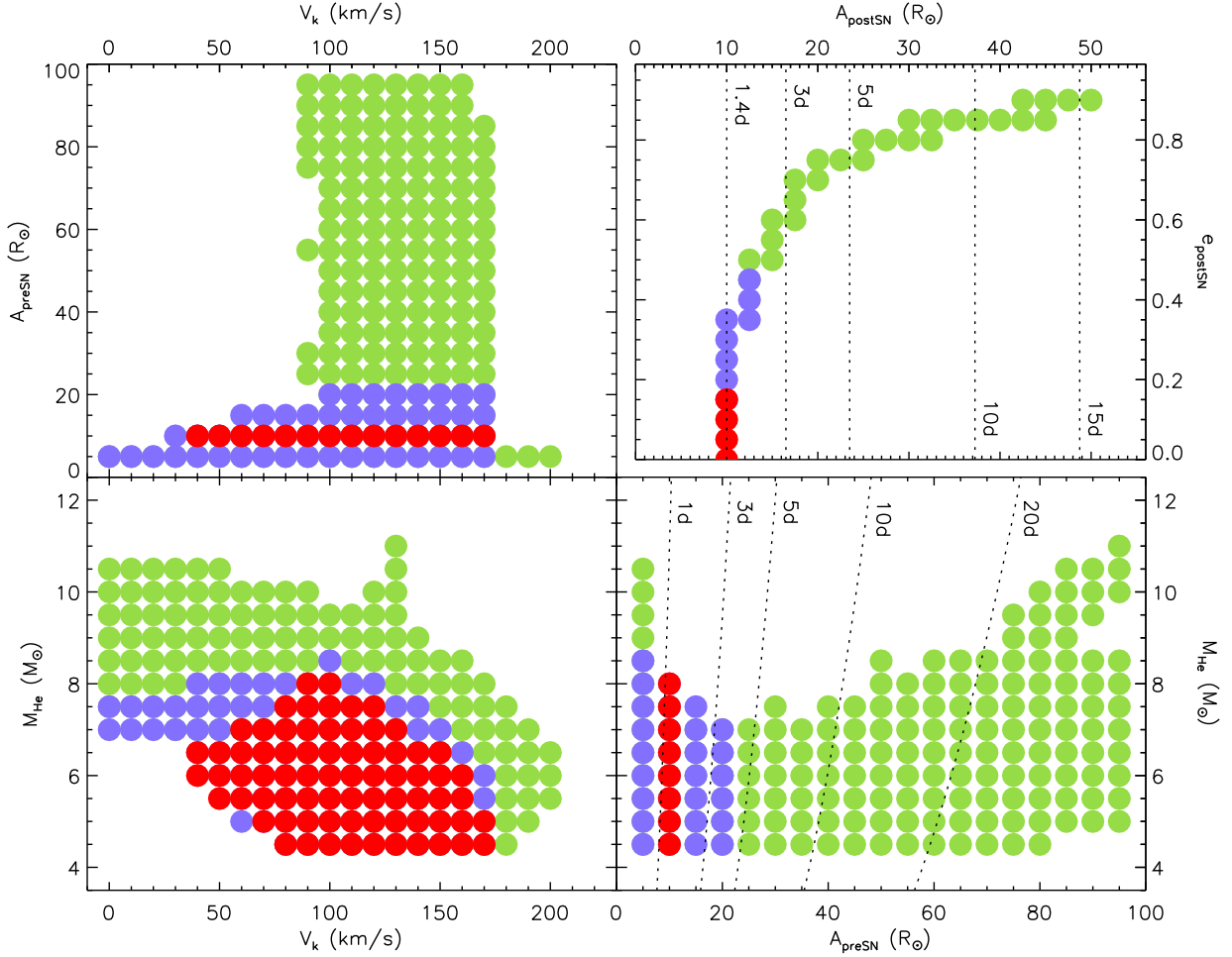}}
\caption{Constraints on the pre-SN helium star mass $M_{\rm He}$, the
magnitude of the kick velocity $V_k$, the pre-SN orbital separation
$A_{\rm preSN}$, and the post-SN orbital separation $A_{\rm postSN}$
and eccentricity $e_{\rm postSN}$ associated with the successful MT
sequence BP\ref{example1} (see Table~\ref{BPseq}). Light grey circles
(green in the electronic edition) correspond to progenitors for which
RLO at periastron occurs more than 200\,Myr earlier than dictated by
the initial age of the donor star in the MT sequence, dark grey
circles (blue in the electronic edition) correspond to progenitors for
which RLO at periastron occurs between 50 and 200\,Myr earlier, and
black circles (red in the electronic edition) correspond to
progenitors for which RLO at periastron occurs within 50\,Myr of the
time dictated by the MT sequence. The black circles (red in the
electronic edition) correspond to our preferred set of solutions. The
dotted lines in the right-hand panels are lines of constant orbital
period.}
\label{eg}
\end{figure*}
\clearpage

To illustrate the derivation of the constraints, we consider the
conservative MT sequence BP\ref{example1} (see Table~\ref{BPseq}) with
initial BH mass $M_{\rm BH} = 4.5\,M_\odot$, initial donor mass
$M_2=2.5\,M_\odot$, and initial orbital period $P_{\rm orb}=1.4$\,days
(note that this is the same sequence as used in Fig.~\ref{orbev}). The
constraints obtained for the mass $M_{\rm He}$ of the BH's helium star
progenitor, the magnitude $V_k$ of the kick velocity imparted to the
BH at birth, the pre-SN orbital separation $A_{\rm preSN}$, and the
post-SN orbital separation $A_{\rm postSN}$ and eccentricity $e_{\rm
postSN}$ are presented in Fig.~\ref{eg}. The tight correlation between
$A_{\rm postSN}$ and $e_{\rm postSN}$ (top tight panel) arises from
matching the post-SN orbital elements to those of the MT sequence at
the onset of RLO. As discussed in Section~\ref{tidaleq}, pairs of
$A_{\rm postSN}$ and $e_{\rm postSN}$ are considered to be
acceptable if, after following the orbital evolution for a time equal
to the age of the donor at RLO onset, the total angular momentum is
equal to that given by the MT sequence within the uncertainties of the
numerical calculations. This condition, however, does not guarantee
that the eccentric orbit can accommodate the BH companion throughout
the time interval between BH formation to RLO. In particular, for
significantly eccentric orbits, RLO at periastron may occur at a time
when the donor star has not yet reached the right evolutionary stage
(see the bottom right panel in Fig.~\ref{orbev} and the associated
discussion). These pairs of $(A_{\rm postSN},e_{\rm postSN})$ will
therefore be valid post-SN orbital parameters only if RLO at
periastron does not significantly affect the star and the orbit. In
view of our limited knowledge of MT in eccentric binaries and its
effects on the orbital elements, we include these solutions in the
presentation of the progenitor constraints, although we consider them
to be less compatible with the initial RLO conditions imposed by the
MT sequence. In order to distinguish between solutions that lead to MT
at periastron at times significantly close to RLO onset in the MT
sequence and those that do not, we separate them into three groups in
Fig.~\ref{eg}: solutions for which RLO at periastron occurs much too
early (i.e.\ more than 200\,Myr) are represented by light grey circles
(green in the electronic edition), solutions for which RLO at
periastron occurs between 50 and 200\,Myr too early are represented by
dark grey circles (blue in the electronic edition), and solutions for
which RLO at periastron occurs within 50\,Myr of the proper time given
by the MT sequence are represented by black circles (red in the
electronic edition). We consider the progenitor constraints associated
with the latter solutions to be the most compatible with the MT
sequence under consideration and therefore refer to them as our
``preferred set'' of solutions.

The constraints on $A_{\rm preSN}$, $A_{\rm postSN}$, and $e_{\rm
postSN}$ are very tight when restricted to progenitors for which RLO
at periastron occurs within $\simeq 50$\,Myr of the donor age imposed
by the MT sequence: $A_{\rm preSN} \simeq 10\,R_\odot$, $A_{\rm
postSN} \simeq 10\,R_\odot$, and $e_{\rm postSN} \la 0.15$. These
ranges broaden somewhat when solutions for which RLO at periastron
occurs between $\simeq 50$ and $\simeq 200\,$Myr too early are
included, but overall the differences with our preferred set of
solutions is not too large. When RLO at periastron occurs more than
$\simeq 200$\,Myr too early, additional highly eccentric post-SN
orbits with large semi-major axes become available. Correspondingly,
the constraints on the pre-SN orbital separation also extend to very
large values. A similar behavior is observed for the constraints on
$M_{\rm He}$ and $V_k$. When RLO at periastron occurs within $\simeq
50$\,Myr of the required donor age, $4.5\,M_\odot \la M_{\rm He} \la
8.0\,M_\odot$ and $40\,{\rm km\,s^{-1}} \la V_k \la 170\,{\rm
km\,s^{-1}}$. For our preferred set of solutions, the formation of the
BH must therefore be accompanied by a natal kick. When solutions for
which RLO at periastron occurs within $\simeq 50-200$\,Myr of the
required donor age are included, the ranges slightly broaden to
$4.5\,M_\odot \la M_{\rm He} \la 8.5\,M_\odot$ and $V_k \la 170\,{\rm
km\,s^{-1}}$. Hence, the derived constraints are fairly robust even
when we allow RLO at periastron to occur up to $\simeq 200$\,Myr too
early. Somewhat larger variations are found when RLO at periastron
occurs more than $\simeq 200$\,Myr too early, but MT for these systems
is likely to significantly affect both the star and the orbit so that
the final parameters after $\simeq 480$\,Myr (the required age of the
donor at the start of RLO) of orbital evolution are probably
incompatible with the initial conditions of the considered MT sequence
and thus these solutions are not favored. Since highly eccentric
post-SN orbital configurations always fall under the latter category,
we do not consider post-SN orbital eccentricities larger than 0.9 in
the derivation of any of the constraints.

To understand the core-collapse event leading to the formation of the
BH, we are mainly interested in the constraints derived for the mass
of the BH's helium star progenitor and the kick velocity that may have
been imparted to the BH at birth. For the remainder of this section,
we therefore restrict ourselves to presenting the constraints derived
for these two quantities. Given that there is no way to distinguish
which of the various successful MT sequences corresponds to the true
progenitor of GRO\,J1655-40, we derive the constraints for each
successful sequence and examine them collectively. The constraints
resulting from our preferred set of solutions for each individual
sequence are summarized in the last two columns of
Tables~\ref{GBOseq}--\ref{BPseq}. 

\clearpage
\begin{figure*}
\resizebox{\hsize}{!}{\includegraphics{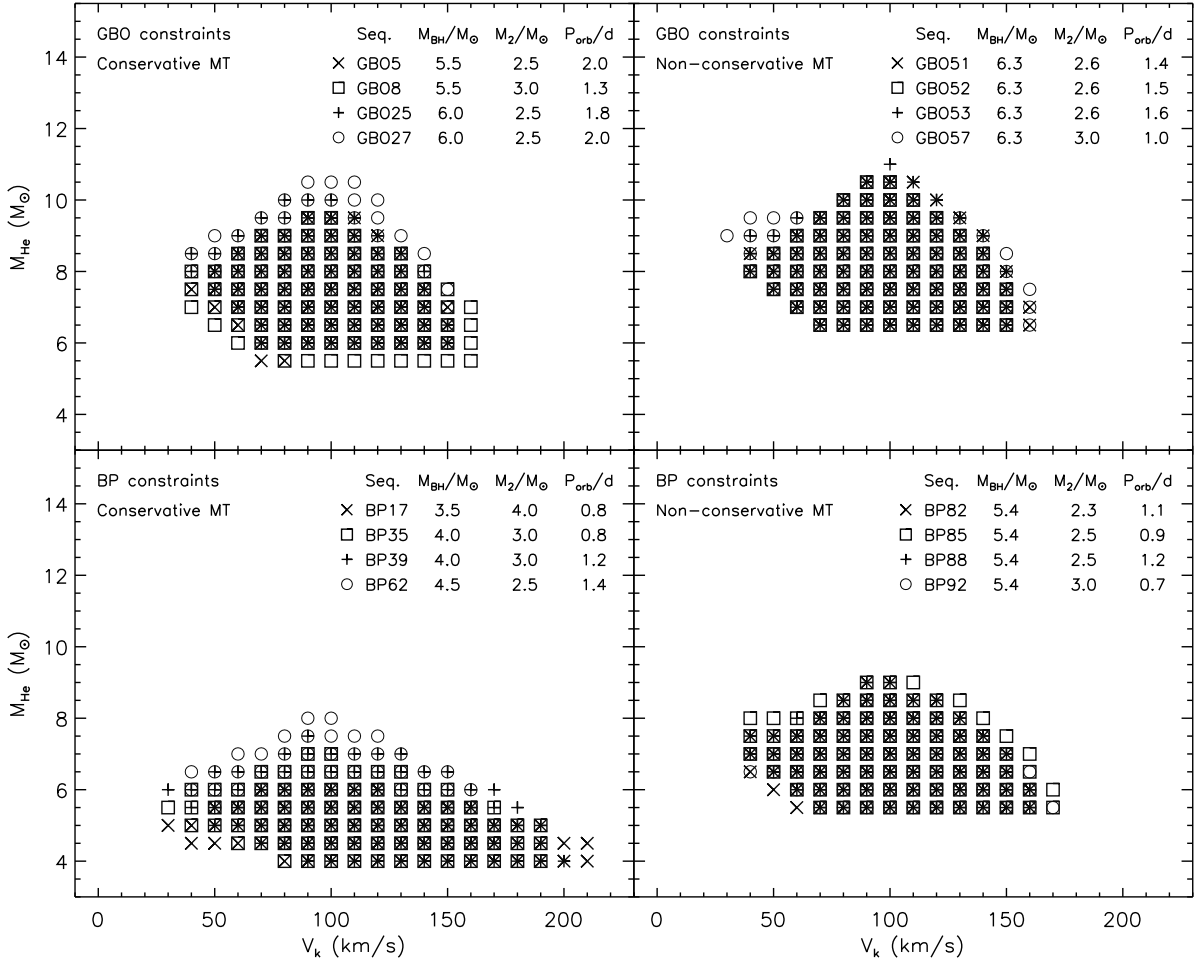}}
\caption{Constraints on the mass $M_{\rm He}$ of the BH's helium star
  progenitor and the magnitude $V_k$ of the kick velocity imparted to
  the BH for some selected successful MT sequences (see
  Tables~\ref{GBOseq}--\ref{BPseq} for details). For clarity, only
  solutions for which RLO at periastron occurs within 50\,Myr of the
  time imposed by the MT sequence are represented.}
\label{overview}
\end{figure*}
\clearpage

In Fig.~\ref{overview}, we first show the constraints on $M_{\rm He}$
and $V_k$ for a selection of MT sequences with initial parameters on
the edges of the regions leading to successful sequences (see
Fig.~\ref{init}) for conservative and non-conservative MT and for GBO
and BP constraints separately. For clarity, only our preferred sets of
solutions are plotted. The variations in the binary properties at the
onset of RLO are seen to yield only small variations in the
constraints for $M_{\rm He}$ and $V_k$. We can therefore be confident
that the derived constraints are not too much affected by the finite
number of successful MT sequences found, as long as we bare in mind
that we outlined the ranges of initial component masses and orbital
periods at the onset of RLO leading to successful sequences to within
a predetermined uncertainty of $0.5\,M_\odot$ in the masses and
0.1\,days in the orbital period.

\clearpage
\begin{figure}
\resizebox{10.0cm}{!}{\includegraphics{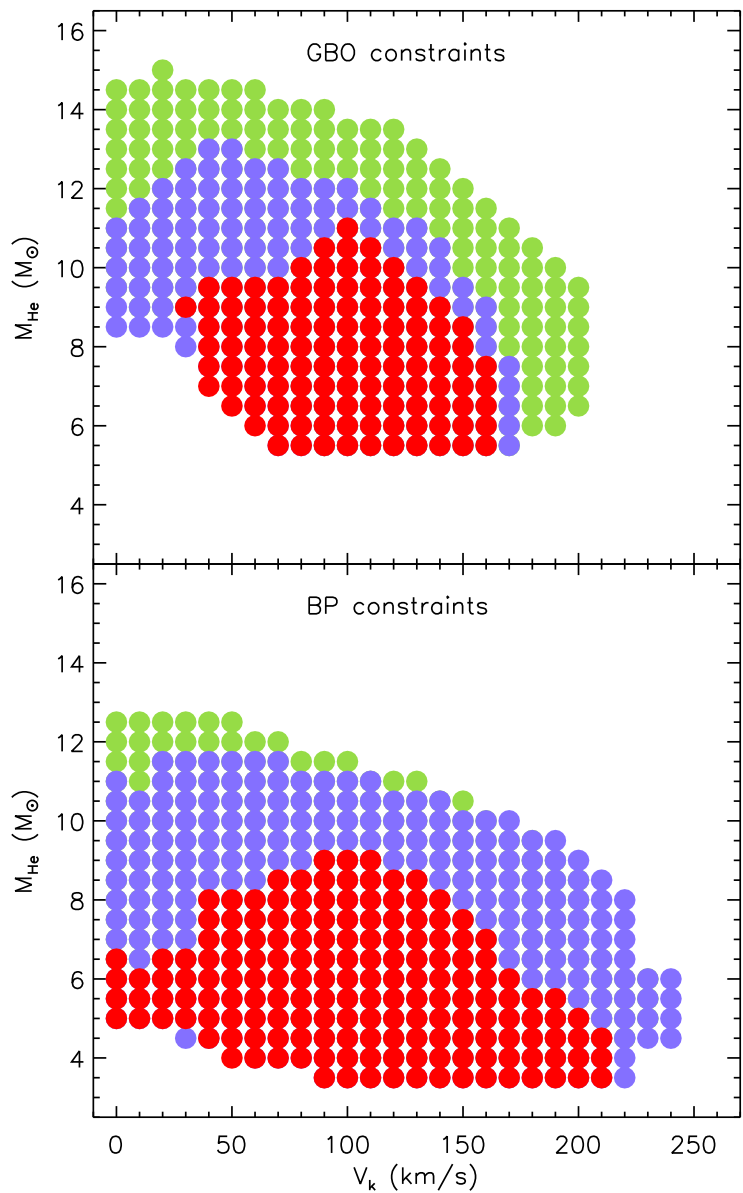}}
\caption{Constraints on the mass $M_{\rm He}$ of the BH's helium star
  progenitor and the magnitude $V_k$ of the kick velocity that may
  have been imparted to the BH obtained by combining the results from
  all successful MT sequences (see Tables~\ref{GBOseq}--\ref{BPseq}
  for sequence details). The different colors have the same meaning as
  in Fig.~\ref{eg}.} 
\label{overview2}
\end{figure}
\clearpage

Since we expect the true MT sequence describing the evolution of
GRO\,J1655-40 from the start of the X-ray phase to its present
configuration to be encompassed by the successful MT sequences, we
overlay the $M_{\rm He}$ and $V_k$ constraints for all the successful
conservative and non-conservative MT sequences found for GBO and BP
system parameters in Fig.~\ref{overview2}. It follows that even the
combination of the extreme cases of fully conservative and fully
non-conservative MT does not significantly relax the constraints on
the progenitor masses and kick velocities with respect those obtained
for individual sequences. For our preferred set of solutions, the
helium star mass and kick velocity are constrained to $5.5\,M_\odot
\la M_{\rm He} \la 11.0\,M_\odot$ and $30\,{\rm km\,s^{-1}} \la V_k
\la 160\,{\rm km\,s^{-1}}$ in the case of GBO parameters, and to
$3.5\,M_\odot \la M_{\rm He} \la 9.0\,M_\odot$ and $0\,{\rm
km\,s^{-1}} \la V_k \la 210\,{\rm km\,s^{-1}}$ in the case of BP
parameters\footnote{It is interesting to note that, for our
preferred set of solutions, the symmetric SN solution in the case of
BP system parameters imparts a kick of $\simeq 50-70\,{\rm
km\,s^{-1}}$ to the binary's center of mass. If all solutions
associated with $V_k = 0\,{\rm km\,s^{-1}}$ are
considered, the mass loss imparts a kick of $\simeq 45-115\,{\rm
km\,s^{-1}}$ to the binary's center of mass, corresponding to the
entire range of admissible post-SN peculiar velocities found from the
Galactic motion calculations presented in Section~\ref{galmotion}.}. On
average, higher progenitor masses furthermore tend to be associated
with lower kick velocities since more mass loss requires less of kick
to achieve the same result.  A similar correlation was found by
Willems et al. (2004) for the progenitor of PSR\,J0737-3039B. 
Overlaying our preferred set of solutions for the other binary
parameters yields $5\,R_\odot \la A_{\rm preSN} \la 15\,R_\odot$,
$10\,R_\odot \la A_{\rm postSN} \la 15\,R_\odot$, and $e_{\rm postSN}
\la 0.3$ for GBO system parameters, and $5\,R_\odot \la A_{\rm preSN}
\la 15\,R_\odot$, $7\,R_\odot \la A_{\rm postSN} \la 10\,R_\odot$, and
$e_{\rm postSN} \la 0.35$ for BP system parameters.

\clearpage
\begin{figure}
\resizebox{10.0cm}{!}{\includegraphics{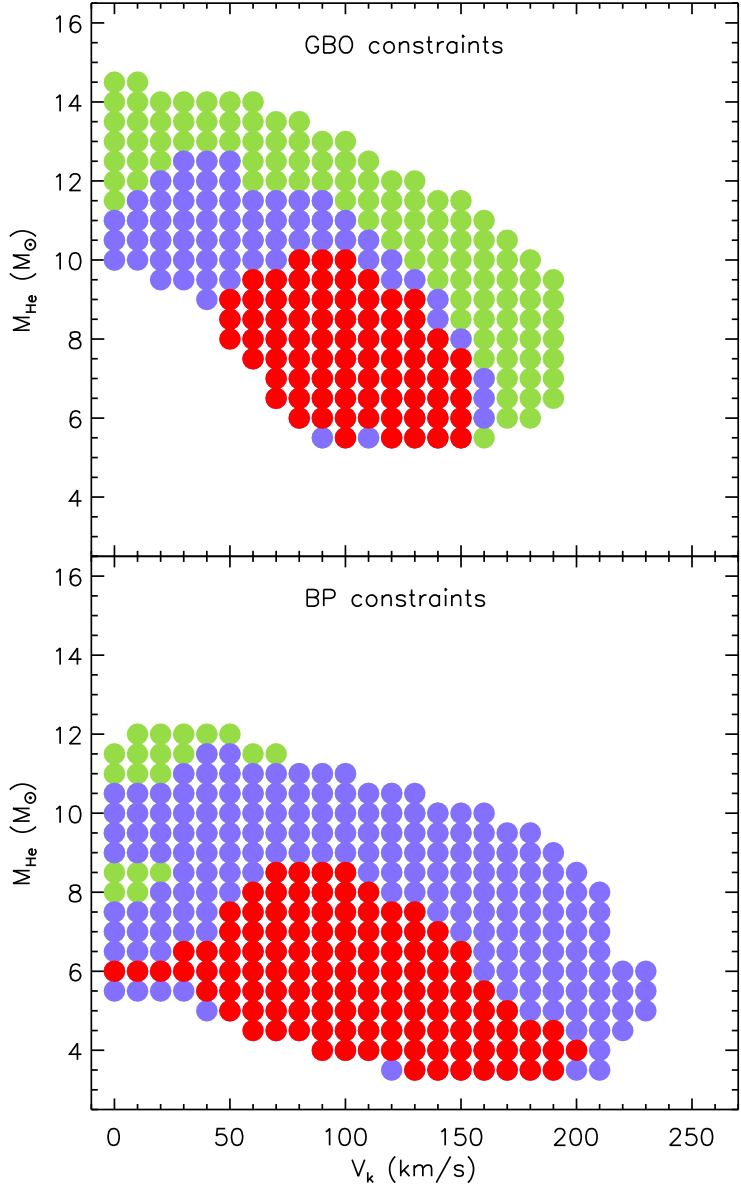}}
\caption{As Fig.~\ref{overview2}, but for a fixed error of $\pm
  10\,{\rm km\,s^{-1}}$ on the post-SN peculiar velocity.} 
\label{overview3}
\end{figure}
\clearpage

As mentioned in Section~\ref{galmotion}, the errors on the post-SN
peculiar velocity, which imposes stringent constraints on $M_{\rm He}$
and $V_k$, were estimated by considering the Galactic orbits of
GRO\,J1655-40 for all possible combinations of the lower and upper
limits on the observed distance and velocity components. Since this
likely overestimates the true errors on the post-SN peculiar velocity,
we reconsidered the derivation of the BH progenitor mass and kick
velocity constraints for a fixed error of $\pm 10\,{\rm km\,s^{-1}}$
on $V_{\rm pec,postSN}$ (represented by the dark grey area in
Fig.~\ref{vcm}). The resulting constraints are shown in
Fig.~\ref{overview3}. For a given helium star mass, the range of
possible kick velocities generally narrows with respect to the range
found in Fig.~\ref{overview2}, although the overall ranges in $M_{\rm
He}$ and $V_k$ do not differ significantly. Are results are therefore
remarkably robust with respect to the uncertainties in the post-SN
peculiar velocity.

We conclude that in the case of GBO parameters, it is likely that the
BH in GRO\,J1655-40 received a kick at birth. Depending on the errors
in the post-SN peculiar velocity, the minimum kick velocity is $\simeq
30$--$50\,{\rm km\,s^{-1}}$. In the case of BP parameters, a kick is
much less of a requirement unless very small errors on the post-SN
peculiar velocity are assumed. The main differences between the
constraints derived for GBO and BP parameters are related to the
higher initial BH masses and the correspondingly higher helium star
masses which reduce the center-of-mass velocity imparted to the binary
by a symmetric SN explosion in the case of GBO parameters.

\section{CORE-COLLAPSE HYDRODYNAMIC SIMULATIONS AND CONSTRAINTS}

To model the collapse of the possible BH progenitors of GRO~J1655-40,
we use the pre-collapse stellar models of Rauscher et al.\ (2003).
Since the binary formed in the galactic disk, we focus on the solar
metallicity models.  We assume the MT phase that revealed the
helium-star occurred late enough in the star's life ($\sim$ case C MT)
that its structure was only slightly modified due to the loss of the
hydrogen envelope, allowing us to simply remove the hydrogen envelopes
from these single-star models to incorporate the effects of close
binary evolution.  One of the major uncertainties in our calculations
is the mass loss.  The high mass-loss rates in current stellar models
are often too high to produce BHs in close binaries.  This excessive
mass-loss is especially extreme in the Wolf-Rayet phase, effectively
making it impossible to make a BH from a star that loses its hydrogen
envelope before the end of helium burning (Wellstein \& Langer 1999).

With current stellar models, we can not make BHs from systems that
lose their hydrogen envelopes in a Case B MT phase.  Even limiting
ourselves to Case C MT, it is difficult to produce helium cores above
11-12\,$M_\odot$.  This is the first constraint that stellar and
core-collapse models can place on the helium star progenitor: If wind
mass-loss rates are correct, $M_{\rm He} \la 11-12$\,$M_\odot$.  This may
be a real constraint, but could also be due to over-estimates of the
mass-loss rates from winds by the stellar evolution community.  To
mimic helium-star BH progenitors of high mass up to 15\,$M_\odot$, we
use the $10^{-4}$ solar metallicity progenitors from Rauscher et al.\
(2003).  Since the metallicity principally effects the mass loss alone
(Heger et al.\ 2003), these models can be used to represent the fate
of solar-metallicity stars with a reduced mass loss.

These stellar models/profiles are mapped into a 1-dimensional
Lagrangian hydrodynamics code (Herant et al.\ 1994; Fryer et al.\
1999).  The equation of state used in the simulations includes a
nuclear equation of state at high densities (Lattimer \& Swesty 1991),
and below $5\times10^{11} {\rm g cm^{-3}}$ an equation of state for
nuclei obeying Boltzmann statistics (Blinnikov, Dunina-Barkovskaya, \&
Nadyozhin 1996, Timmes \& Arnett 1999).  Both in the high density
equation regime and in the low-density regime above temperatures of
$5\times10^{9}$\,K, abundances are determined assuming nuclear
statistical equilibrium. Between $2\times10^{8}$ and $5\times10^{9}$,
the abundances are evolved using a 14-element nuclear network (Benz,
Hills, \& Thielemann 1989) which yields reasonable approximations for
the nuclear energy generation rate (Timmes, Hoffman \& Woosley 2000).

\clearpage
\begin{deluxetable}{lcccccc}
%\tablewidth{18pc} 
\tablecolumns{7}
\tablecaption{Core-Collapse Models \label{ccmod}}

\tablehead{ \colhead{Progenitor} & \colhead{$f_\nu$\tablenotemark{a}}
& \colhead{Remnant} & \colhead{Energy$_{\rm 3000\,km}$} &
\colhead{K.E.$_{\rm breakout}$\tablenotemark{b}} & 
\colhead{t$_{\rm remnant}$} &
\colhead{V$_{\rm remnant}$\tablenotemark{c}} \\ 
\colhead{Mass ($M_\odot$)} &
\colhead{} & \colhead{Mass ($M_\odot$)} & \colhead{($10^{51}$\,erg)} &
\colhead{($10^{51}$\,erg)} & \colhead{$10^6$\,yr} &
\colhead{V$_{51}$}}

\startdata

6.2 & 0.9 & 6.2 & 0.0 & 0. & 0. & 0. \\
6.2 & 1 & 4.59 & 0.9 & 0.068 & 1.0 & 0.08 \\
6.2 & 1.1 & 2.39 & 1.4 & 0.19 & 1.4 & 0.2 \\
8.2 & 1.0 & 6.33 & 0.8 & 0.083 & 1.0 & 0.09 \\
8.2 & 1.1 & 5.42 & 1.0 & 0.17 & 1.3 & 0.18 \\
8.2 & 1.2 & 3.16 & 1.6 & 0.32 & 1.6 & 0.33 \\
10.0 & 1.0 & 7.14 & 0.8 & 0.16 & 1.3 & 0.17 \\
10.0 & 1.06 & 4.16 & 1.6 & 0.64 & 2.0 & 0.65 \\
10.0 & 1.12 & 2.26 & 2.0 & 1.3 & 2.5 & 1.3 \\
14.9 & 0.92 & 14.9 & 0. & 0. & 0. & 0. \\
14.9 & 1.00 & 5.34 & 7.1 & 2.26 & 3.0 & 2.2 \\
14.9 & 1.04 & 2.23 & 12.0 & 7.7 & 4.4 & 7.1 \\

\enddata

\tablenotetext{a}{$f_\nu=1$ corresponds to the lowest factor 
used that still produces an explosion.  All other values are 
relative to this quantity.}
\tablenotetext{b}{Kinetic energy}
\tablenotetext{c}{The effective volume covered by the SN
remnant before it merges with the interstellar medium as a fraction 
of the volume encompassed by a $10^{51}$\,erg explosion.}

\end{deluxetable}
\clearpage

Neutrino transport is mediated using an explicit flux-limited
algorithm (Herant et al.\ 1994).  Above a radius determined by the
neutrino mean free path, the neutrinos are transported in the free
streaming limit, escaping the star instantly.  For most models
(e.g., Fryer et al.\ 1999), this radius is chosen such that less than
3-5\% of the neutrinos in the free streaming limit interact with the
matter as it leaves the star.  In the models presented here, the
neutrinos in this free streaming limit are increased by a factor
$f_\nu$ and the limit is set to force roughly 10\% of the neutrinos in
the free-streaming limit to interact with matter.  This effectively
raises the neutrino energy in the ``gain'' region where neutrino
heating deposits energy into the stellar material.  By increasing
$f_\nu$, we increase the neutrino heating and can drive a SN explosion
(Table~\ref{ccmod} lists the suite of these simulations). 

The high sensitivity of remnant mass on our fudge factor $f_\nu$
demonstrates just how delicate the balance between success and failure
is for the neutrino-driven mechanism.  This delicate balance is one
reason why core-collapse theorists have struggled to get consistent
explosions in their simulations of these explosions (getting all of
the necessary physics accurate to the 10\% level is a daunting task).
Although a robust mechanism may actually exist where 10\% effects are
not important, it is important to understand that the actual
progenitors of core-collapse (massive stars with masses above
8--9\,$M_\odot$) are not too dissimilar, yet nature produces a range 
of remnant masses.   It appears that nature too is sensitive to 
small changes in the details.

A massive star collapses until it reaches nuclear densities.  A
proto-NS is formed and the explosion occurs if the neutrinos from the
cooling NS can provide enough energy to drive off the rest of the
star.  As the explosion is launched, matter expands off the NS
surface, and the cells are split to retain resolution with our
Lagrangian code near the NS surface.  After $\sim$1\,s the explosion
shock is well-developed ($\sim 3000$\,km).  At this point, the NS is
cut out of the simulation and modelled as a gravitational source with
a hard boundary.  The explosion is then followed until the shock
reaches the edge of the surface of the helium star.  At this point,
the energy in the exploding star is primarily in kinetic energy and we
can determine the rough SN explosion energy of these objects (see
Table~\ref{ccmod}).  We also determine the BH mass by assuming all
material with velocities less than the escape velocity will ultimately
fall back onto the NS (see Table~\ref{ccmod}).  For all cases, we can
produce a BH with a mass lying somewhere between $\sim
4.0$--$6.5\,M_\odot$, but the resulting energies for these BHs lies
between $\sim 0-2 \times 10^{51}$\,erg.

Our understanding of core-collapse theory can also be used to
constrain the BH progenitors.  Fryer (1999) found that, using a
multi-dimensional core-collapse code, massive stars above
$\sim$20\,$M_\odot$ achieved explosion energy (when the shock was near
1000-3000\,km) of roughly $0.6\times10^{51}$\,erg.  There are a number
of uncertainties in these ``first-principle'' collapse calculations
and this answer can change considerably with more accurate codes.  But
if we take these simulations at face value and assume that the energy
is, within a factor of 2, accurate, we can also rule out some
progenitors.  From Table~\ref{ccmod}, we see that only progenitors
below 10\,$M_\odot$ can obtain the appropriate BH mass with less than
about $1.2\times10^{51}$\,erg of explosion energy when the shock is
launched.  This constraint (the second from stellar and core-collapse
models) is strongest for small BH masses.

\clearpage

\begin{figure*}
\resizebox{15.0cm}{!}{\includegraphics{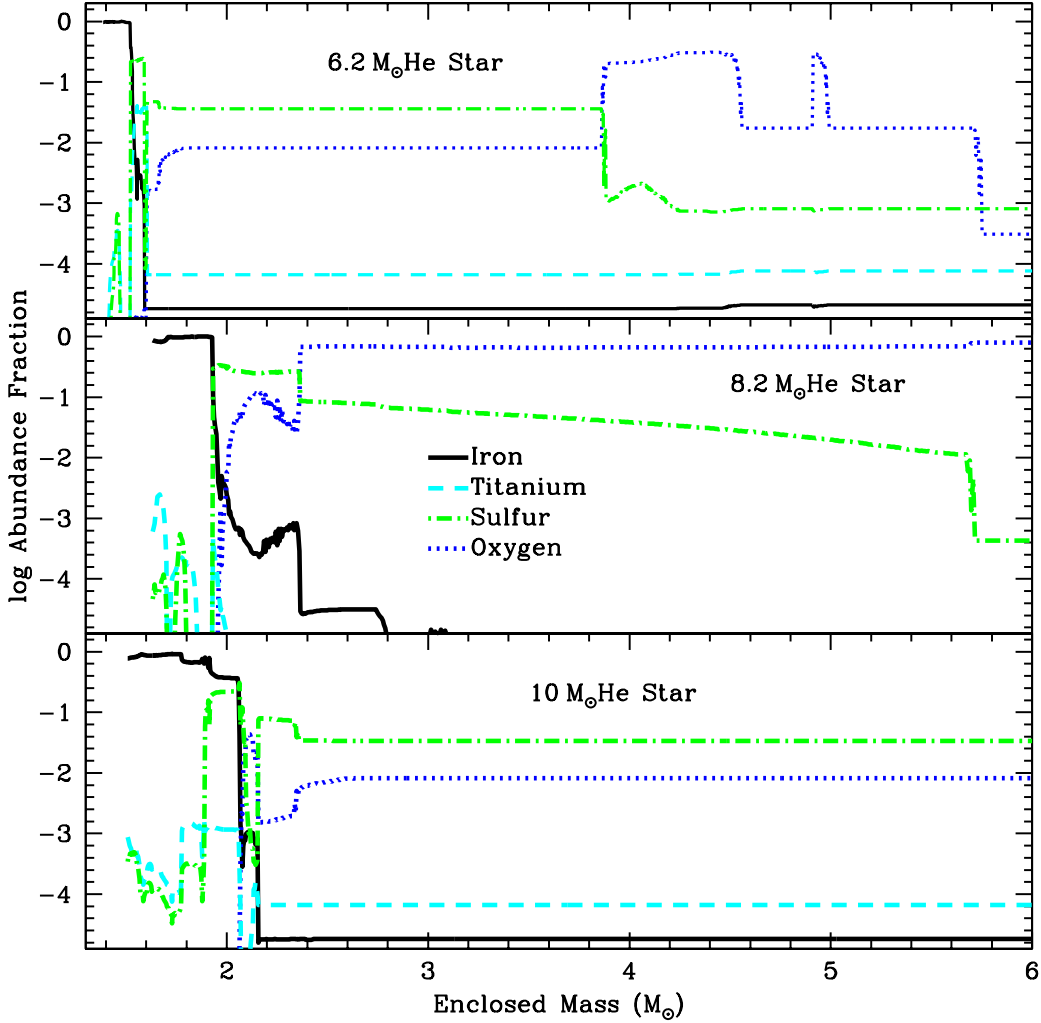}}
\caption{Abundance fractions of iron (solid line), titanium (dashed
  line), sulfur (dot-dashed line), and oxygen (dotted line) versus
  enclosed mass for three models: $M_{\rm He}=6.2\,M_\odot$, $f_\nu=1$
  (top), $M_{\rm He}=8.2\,M_\odot$, $f_\nu=1.1$ (middle), and $M_{\rm
  He}=10.0\,M_\odot$, $f_\nu=1.06$ (bottom).  In these explosions,
  roughly the inner 4-5\,$M_\odot$ of material will fall back onto the
  remnant, forming a BH.  Without mixing, none of the explosions
  will produce iron or titanium far enough out to enrich the
  companion.  Indeed, using estimates from Hungerford et al. (2003),
  even the mixing that occurs in symmetric 3-dimensional simulations
  is not sufficient to lead to iron and titanium enrichment of the
  companion.  Only asymmetric explosions can produce enough mixing. 
  The level of asymmetry required depends primarily on how far out the
  iron and titanium are formed.  Notice that the iron and titanium is
  formed nearly twice as far out in mass coordinate for the
  10\,$M_\odot$ progenitor compared to the 6.2\,$M_\odot$ progenitor.
  The level of asymmetry to produce the necessary enrichment is much
  higher for the 6.2\,$M_\odot$ progenitor when compared to the
  10\,$M_\odot$ progenitor.}
\label{fig:abun}
\end{figure*}
\clearpage

We can also use the nucleosynthetic yields to constrain the models.
As noted above, the 14 isotope nuclear network used in the
hydrodynamics simulations is used only to determine energetics. To
obtain more detailed abundance information, we post-processed the
thermodynamic trajectories of our Lagrangian particles with a modern
500 isotope network (e.g., Timmes 1999). For simplicity we assumed an
initial electron fraction of 0.5.  The iron, titanium, sulfur and
oxygen yields as a function of enclosed mass for three representative
models are presented in Fig.  \ref{fig:abun}.  These three models:
$M_{\rm He}=6.2\,M_\odot$, $f_\nu=1$, $M_{\rm He}=8.2\,M_\odot$,
$f_\nu=1.1$, $M_{\rm He}=10\,M_\odot$, $f_\nu=1.06$, each produce BHs
with masses above 4\,$M_\odot$.  Oxygen and even some sulfur is formed
well out in the star, but titanium and iron are formed in the most
intense burning at the base of the explosion.  However, note that the
more massive stars produce iron and titanium further out than the
low-mass stars.  This occurs because the more massive stars require
stronger explosions to produce BHs of comparable mass to the lower
mass stars.  The stronger explosions produce intensive nuclear burning
further out in the star, ultimately producing iron further out in the
star as well.

Do such abundance profiles provide additional, useful information?  We
often use the nucleosynthetic signature from supernovae to constrain
SN progenitors.  However, the yields from these stars must mix out
extensively in the explosion just to get ejected.  These BH forming
stars will not dominate the nuclear yields of the Galaxy.  In
addition, due to their weak explosions, the nucleosynthetic yields
from these explosions are not likely to mix extensively into the
interstellar medium.  The timescale ($t_{\rm max}$) for the SN shock
to once again become part of the interstellar medium (when the shock
velocity drops below the sound speed of the ambient medium) is roughly
(McKee \& Hollenbach 1980):
\begin{equation}
t_{\rm max}=2.3 \times 10^6 E^{0.32}_{51} c_{\rm s 6}^{-1.45}
n_0^{-0.36} {\rm yr,}
\end{equation}
where $E_{51}$ is the explosion energy in units of $10^{51}$\,erg; and
$c_{\rm s 6} \approx 1 $ in units of $10^6$cm\,s$^{-1}$ and $n_0
\approx {\rm pc^{-3}}$ are, respectively, the sound speed and particle
density of the ambient medium.  The corresponding maximum extent
($r_{\rm max}$) of the SN shock is:
\begin{equation}
r_{\rm max}=76 E^{0.32}_{51} c_{\rm s 6}^{-0.45}
n_0^{-0.36} {\rm pc.}
\end{equation}
The timescales and effective volumes (as a fraction of a normal
$10^{51}$\, erg explosion) of the remnants of these BHs are given in
Table~\ref{ccmod}.  With such short-lived, small remnants, it is
unlikely that we will be able to constrain these stars by Galactic or
remnant abundances.

However, Israelian et al. (1999) observed the companion of GRO
J1655-40 during a quiescent X-ray phase using the high-resolution
spectrograph at Keck.  At visible wavelengths, they found that the
abundances of O, Mg, S, Si, and Ti were 6-10 times larger than solar.
They also found that [Fe/H] = 0.1 $\pm$ 0.2, i.e. almost solar.  The
most straightforward and logical way of explaining these
overabundances, and the one proposed by Israelian et al., is that
these elements were produced during explosive oxygen and silicon
burning during the BH formation and have enriched the companion.
But to cause this enrichment, these elements must make it to the
companion star.  For the iron and titanium, this can be quite a
difficult task.  At the time of explosion the companion was roughly
$5-15\,R_\odot$ from the center of the collapsing star.   But our
1-dimensional explosion models find that the iron and titanium does
not make it beyond 0.05\,$R_\odot$ before falling back onto the
compact remnant.  Even in 3-dimensional simulations, the amount
of mixing found in symmetric models does not place the iron or
titanium beyond 0.5\,$R_\odot$ before these elements fall back
onto the compact remnant. 
The extensive mixing required to eject these elements
requires large asymmetries in the explosion (Hungerford et al. 2003;
Hungerford et al. 2004). Hence, if the enrichment of the companion of
GRO J1655-40 is arising from the explosion forming the BH, its SN must
have been asymmetric!  The amount of asymmetry reaches a minimum with
the lower mass progenitors (especially with massive BHs).  For
example, in the extreme case of a BH mass of 6.2\,$M_\odot$, the
6.2\,$M_\odot$ progenitor will not, no matter what the asymmetry, be
able to enrich the companion.  Even with smaller BH masses, it is
likely that the helium progenitor for this system is more massive
($\ga 8\,M_\odot$).  This is our third, and final, constraint on the
helium star mass.  

In principal, the exact abundances of the companion star could be used
to determine the best-fitting BH progenitor (see Podsiadlowski et
al. 2002b).  However, the uncertainties in the level of asymmetry and
the uncertainties of the effects of asymmetry on the actual nuclear
yield make it difficult to produce any accurate quantitative results.
These uncertainties must then be folded into uncertainties in the
accretion onto the companion.  The study of asymmetry-induced mixing
is still in its infancy.  When it is better understood, the actual
composition of the companion star can be used to constrain the
progenitors as well.

Although there exist a number of uncertainties in stellar evolution
and core-collapse calculations, these models can place additional
constraints on the BH progenitor for GRO\,J1655-40:
\begin{itemize}
\item{\bf I: $M_{\rm He} \la 11-12\,{\rm M}_\odot$ - Stellar Winds.} With
  current mass-loss rates for winds, helium cores above this mass are
  simply not produced.  The limitation of this constraint lies in the
  uncertainties in stellar winds. Given that BH masses for some other
  systems have been estimated to be as high as $18\,M_\odot$
  (McClintock \& Remillard 2004, and references therein), we do not
  consider this constraint to be very stringent.
\item{\bf II: $M_{\rm He} \la 10\,{\rm M}_\odot$ - Supernova Energetics.}
  If explosion energies for these massive stars do decrease with
  increasing stellar mass, it is impossible to get low-mass BHs with
  more massive helium stars.  This constraint is strongest for the
  low-end of the BH mass range allowed for GRO J1655-40.  This
  constraint is limited by uncertainties in the core-collapse
  mechanism.
\item{\bf III: $M_{\rm He} \ga 8\,{\rm M}_\odot$ - Nucleosynthetic
  Yields.} If the high abundances of the companion to GRO J1655-40 are
  due to ejecta enrichment, large asymmetries are required to mix out
  the iron and titanium so that it is ejected.  Prohibitively large
  asymmetries are required for low-mass progenitors.  This constraint
  is strongest for BH masses at the high-end of the allowed range for
  GRO J1655-40.  The dominant limitation in this constraint arises
  from the uncertainty of the true enrichment in the companion of GRO
  J1655-40.
\end{itemize}
In addition to constraints on the progenitor mass, our calculations
also show how observations of GRO J1655-40 can constrain SN
explosions.  If the anomalous abundances in the companion to GRO
J1655-40 truly arise from enrichment from the BH-forming SN, then this
object provides some of the best evidence not only for the occurrence
of a SN explosion, but also for asymmetries in the explosions
(Israelian et al. 1999, Podsiadlowski et al. 2002b).  A
better constraint on the progenitor (as well as a better understanding
of structure of the asymmetries) could significantly constrain the
magnitude of the asymmetry in this explosion.

\section{DISCUSSION AND CONCLUSIONS}

In this paper, we initiated a comprehensive study on the formation and
evolution of Galactic XRBs. Based on the current binary properties and
the current position and motion of observed XRBs, the kinematic and
evolutionary history is traced back in time and constraints are
derived for the pre-SN binary progenitor and for the kick velocity
that may have been imparted to the compact object at birth. The
analysis consists of the following elements: (i) MT calculations to
model the ongoing X-ray phase, (ii) orbital evolution due to tides and
gravitational radiation between the compact object formation time and
the onset of RLO, (iii) motion of the system in the Galaxy after the
formation of the compact object, (iv) binary orbital dynamics at the
time of core collapse, and (v) hydrodynamic modeling of the actual
core collapse event.

As a first application, we constrained the progenitor properties and
the formation of the BH in the soft X-ray transient GRO\,J1655-40. For
this purpose, we calculated a large number of MT sequences to map the
currently observed binary properties (component masses, position of
the donor in the H-R diagram, orbital period, and transient behavior)
to those at the onset of the MT phase. Uncertainties related to the
accretion of matter by the BH are incorporated by considering both
conservative (for sub-Eddington rates) and non-conservative MT. For
the current system configuration, we considered both the parameters
derived by GBO and those derived by BP (see Table~\ref{1655param} for
details).  In the case of GBO parameters, successful MT sequences able
to reproduce all currently observed system properties were found for
initial BH masses $M_{\rm BH} \simeq 5.5$--$6.3\,M_\odot$, initial
donor star masses $M_2 \simeq 2.5$--$3.0\,M_\odot$, and initial
orbital periods $P_{\rm orb} \simeq 1.0$--$2.0$\,days. In the case of
BP parameters, successful sequences were found for initial BH masses
$M_{\rm BH} \simeq 3.5$--$5.4\,M_\odot$, initial donor star masses
$M_2 \simeq 2.3$--$4.0\,M_\odot$, and initial orbital periods $P_{\rm
orb} \simeq 0.7$--$1.5$\,days. For both cases, all successful
sequences have a MS donor star at the onset of RLO.  A graphical
representation of the initial donor masses and orbital periods leading
to successful MT sequences is presented in Fig.~\ref{init}. Properties
of the sequences are given in Tables~\ref{GBOseq}--\ref{BPseq}.

The results of the MT calculations are in good agreement with previous
investigations. Reg\H{o}s, Tout, \& Wickramasinghe (1998) showed that
the present-day properties of GRO\,J1655-40 (as determined by Orosz \&
Bailyn 1997) could be reproduced by a binary initially consisting of a
$2.75\,M_\odot$ MS star orbiting a $6.6\,M_\odot$ BH with a period of
1.935\,days. These initial parameters are fairly close to those of the
successful MT sequence GBO\ref{rtw} in Table~\ref{GBOseq}, except for
the higher initial BH mass adopted by Reg\H{o}s et al. (1998). This
small difference can be attributed to the higher current BH mass of
$7.02 \pm 0.22\,M_\odot$ estimated by Orosz \& Bailyn (1997) as well
as to some different input physics in the stellar evolution codes,
such as, e.g., the amount of convective overshooting.

Kolb et al. (1997) and Kolb (1998), on the other hand, suggested that
GRO\,J1655-40 may be in a very short-lived evolutionary phase in the
Hertzsprung gap during which the expansion of the donor's radius
temporarily halts or even reverses. As a consequence, the MT rate
drops below the high rates typically expected for Hertzsprung-gap
donor stars, allowing for a short phase of transient behavior. In view
of the small size of the relevant region in the Hertzsprung gap (see,
e.g., Fig. 2 in Kolb et al. 1997) and the associated short
evolutionary time scales, progenitors which start RLO while the donor
is still on the MS seem statistically more favorable. We therefore did
not attempt to reproduce the small transient strip in the Hertzsprung
gap found by Kolb et al. (1997) and Kolb (1998). In their
investigation, Kolb et al. (1997) also adopted the BH mass derived by
Orosz \& Bailyn (1997), which is lower than that derived by both GBO
($M_{\rm BH}=6.3 \pm 0.5\,M_\odot$) and BP ($M_{\rm BH}=5.4 \pm
0.3\,M_\odot$). Since, for a given donor mass and orbital period, the
MT rate tends to increase with decreasing BH mass, the small transient
strip in the Hertzsprung gap may disappear when the more recent BH
mass estimates are adopted.

BP, finally, found that, using their estimates for the system
parameters, both case~A and early case~B MT sequences are able to
reproduce the observed system properties, but that case~A sequences
generally provided more satisfactory fits. Their example sequence with
initial parameters $M_{\rm BH}=4.1\,M_\odot$, $M_2=2.5\,M_\odot$, and
$P_{\rm orb}=0.8$\,days is close to our sequence BP\ref{bpeg} (see
Table~\ref{BPseq}), although the latter has a somewhat higher initial
donor mass, possibly due to differences in the adopted convective
overshooting.

A crucial element emerging from the MT calculations is the age of the
donor star when all observational constraints are satisfied (i.e.\ the
donor's {\em current} age). By using this age as an estimate for the
time expired since the BH formation, we determined the possible birth
sites of the BH by following the motion of the system in the Galaxy
backwards in time. It follows that GRO\,J1655-40 was born between 3
and 7\,kpc from the Galactic center, and within 200\,pc from the
Galactic plane. The post-SN peculiar velocity at the birth site can be
anywhere between $\simeq 45$ and $\simeq 115\,{\rm km\,s^{-1}}$. We
use this post-SN peculiar velocity to constrain the amount of mass
lost in the SN explosion and the kick velocity that may have been
imparted to the BH. This is a major difference with previous
investigations on the kinematics of GRO\,J1655-40 which used the {\em
present-day} radial velocity (after correcting for the motion of the
Sun and differential Galactic rotation, $V_r=-114 \pm 19\,{\rm
km\,s^{-1}}$; Brandt et al. 1995) as a lower limit to the binary's
center-of-mass velocity (see Brandt et al. 1995, Nelemans et al. 1999,
Fryer \& Kalogera 2001).

The MT sequences also yield the age of the donor at the onset of
RLO. Knowledge of this age allows us to link the post-SN orbital
parameters to those at the onset of RLO by numerically integrating the
system of differential equations governing the orbital evolution due
to tides and gravitational radiation. Post-SN orbital semi-major axes
and eccentricities compatible with successful MT sequences range from
$A_{\rm postSN}=5\,R_\odot$ to $A_{\rm postSN}=15\,R_\odot$ and from
$e_{\rm postSN}=0$ to $e_{\rm postSN}=0.35$.

The solution of the equations describing the conservation of orbital
energy and angular momentum during compact object formation allows us
to derive constraints on the pre-SN orbital separation, on the mass of
the BH's helium star progenitor, and on the magnitude of the kick
velocity that may have been imparted to the BH at birth. Despite the
large number of successful MT sequences found, the progenitor
constraints turn out to be fairly robust to the initial parameters of
the sequences at the onset of RLO (see Fig.~\ref{overview}). Combining
the constraints obtained for all successful MT sequences for both
conservative and non-conservative MT, yields $M_{\rm He} \simeq
5.5$--$11.0\,M_\odot$ and $V_k \simeq 30$--$160\,{\rm km\,s^{-1}}$ for
GBO system parameters and $M_{\rm He} \simeq 3.5$--$9.0\,M_\odot$ and
$V_k \simeq 0$--$210\,{\rm km\,s^{-1}}$ for BP system parameters (see
Fig.~\ref{overview2}). It is clear that symmetric SN explosions lie at
the edge of our solution ranges and that BH kicks of at least a few
tens of km\,s$^{-1}$ are favored by the majority of the successful MT
sequences (see Tables~\ref{GBOseq}--\ref{BPseq}).

Hence, although some of the constraints associated with individual MT
sequences require a kick to be imparted to the BH at birth (see
Figs.~\ref{eg}--\ref{overview3} and Tables~\ref{GBOseq}--\ref{BPseq}),
without any additional information, the still remaining uncertainties
in the present-day system parameters at this stage do not allow us to
exclude that the BH in GRO\,J1655-40 may have been formed through a
symmetric SN explosion. In the final step of our investigation, we
therefore model the core collapse using a 1D Lagrangian hydrodynamics
code. From the energetics of the SN explosion, it follows that the
mass of the BH's helium star progenitor must be larger than
$8\,M_\odot$. In addition, if the observed overabundances of heavy
elements in the BH's companion are to be attributed to pollution from
SN ejecta, the nucleosynthesis yields imply a maximum helium-star mass
of $10\,M_\odot$. Combining this with the constraints derived from the
MT sequences, orbital dynamics, and motion in the Galaxy gives
$30\,{\rm km\,s^{-1}} \la V_k \la 150\,{\rm km\,s^{-1}}$ in the case
of GBO parameters and $40\,{\rm km\,s^{-1}} \la V_k \la 140\,{\rm
km\,s^{-1}}$ in the case of BP parameters.
  
Our final set of constraints on the progenitor are much different than
those obtained by Podsiadlowski et al. (2002b) who predict the likely 
progenitor to lie between $10-16\,M_\odot$.  But these differences
are not due to different estimates of a given constraint, but the
application of different constraints.  Podsiadlowski et al. (2002b)
focus their attention on the enrichment of the companion.  As with our
enrichment constraint (constraint III from \S 9), they find that more
massive stars are better suited to explain the nucleosynthetic
enrichment in the companion.  Our constraint is slightly weaker, based
on the simulations of Hungerford et al. (2003, 2004) which suggest that
asymmetries can lead to very extended mixing.  We have an additional
limit based on both our mass-loss and supernova explosion model
constraints (I and II from \S 9) that push for lower-mass progenitors.  
Since Podsiadlowski et al. (2002b) don't consider these constraints, 
they do not have this upper limit in their paper.

The question of whether or not a natal kick was imparted to the BH in
GRO\,J1655-40 was addressed previously by Brandt et al. (1995),
Nelemans et al. (1999), and Fryer \& Kalogera (2001). All three of
these investigations used a lower limit on the present-day
center-of-mass velocity given by the {\em current} radial velocity
($V_r=-114 \pm 19\,{\rm km\,s^{-1}}$) instead of the actual post-SN
peculiar velocity to constrain the mass loss and kick magnitude
associated with the BH's formation. As illustrated in Fig.~\ref{vcm},
this neglects changes in the system's center-of-mass velocity
resulting from its acceleration in the Galactic potential. We here
find that, even though following the Galactic motion of the system
backward in time yields post-SN peculiar velocities all the way down
to $\simeq 45\,{\rm km\,s^{-1}}$, the additional constraints on the
binary properties make the possibility of a symmetric SN explosion
only marginally acceptable solutions.

Jonker \& Nelemans (2004) recently reconsidered the vertical
distribution of low-mass XRBs with respect to the Galactic plane and
found no significant difference between NS and BH systems, suggesting
that BHs may be subjected to natal kicks as well. The authors
furthermore noted that the distances to Galactic soft X-ray transients
may be systematically underestimated. If this is truly so, revised
distance estimates may have important consequences for the
determination of quiescent X-ray luminosities, peak outburst
luminosities, BH masses, and Galactic distribution of BH
XRBs. Clearly, if future measurements significantly revise any of the
observational constraints, the analysis presented in this paper should
be repeated to account for the most up-to-date knowledge of
GRO\,J1655-40's current system properties.

In subsequent analyses, we intend to apply the procedure outlined
above to the soft X-ray transient XTE\,1118+480, the high-mass XRBs
Cyg X-1, LS\,5039, LSI\,$+61^\circ\,303$, Vela\,X-1, 4U1700-37, and the
NS low-mass XRB Sco\,X-1. By examining both NS and BH systems and both
RLO and stellar wind induced MT, we hope to unravel the systematic
dependencies between the masses of newly formed compact objects and
their immediate pre-SN progenitors, the mass lost at core collapse,
and the possible kick velocity magnitude imparted to the compact
object at birth. We expect that such constraints will help us
understand better the physical origin of asymmetries in the collapse
of massive stars.

\acknowledgments 

We are indebted to Laura Blecha for sharing the code used to follow
the motion of GRO\,J1655-40 in the Galactic potential, and to Jon
Miller, Jerome Orosz, Jeffrey McClintock, Klaus Schenker, and Gijs
Nelemans for useful and stimulating discussions. This work is
supported by a David and Lucile Packard Foundation Fellowship in
Science and Engineering grant and made extensive use of NASA's
Astrophysics Data System Bibliographic Services.

\end{document}